\documentstyle[epsfig,aps]{revtex}
\begin{document}
\draft
\title{Super-radiant light scattering from trapped Bose Einstein condensates}
\author{\"{O}zg\"{u}r E. M\"{u}stecapl{\i}o\u{g}lu and L. You}
\address{School of Physics, Georgia Institute of Technology,
Atlanta, Georgia 30332-0430}
\date{\today}
\maketitle

\begin{abstract}
We propose a new formulation for atomic side mode dynamics from
super-radiant light scattering of trapped atoms. A detailed
analysis of the recently observed super-radiant light scattering
from trapped bose gases [S. Inouye {\it et al.}, Science {\bf
285}, 571 (1999)] is presented. We find that scattered light
intensity can exhibit both oscillatory and exponential growth
behaviors depending on densities, pump pulse characteristics,
temperatures, and geometric shapes of trapped gas samples. The
total photon scattering rate as well as the accompanied matter
wave amplification depends explicitly on atom number fluctuations
in the condensate. Our formulation allows for natural and
transparent interpretations of subtle features in the MIT data,
and provides numerical simulations in good agreement with all
aspects of the experimental observations.
\end{abstract}


\narrowtext

\section{INTRODUCTION}
The successful discovery of Bose-Einstein condensation (BEC) in
dilute trapped atoms \cite{anderson} provided significant momentum
for research into quantum degenerate gases.
In analogy with laser theory, condensation results in a
coherent matter wave field, which has since been identified \cite{jila},
and several important optical analogous effects including four-wave
mixing \cite{deng}, superradiance \cite{inouye}, coherent matter
wave growth \cite{inouye2} were demonstrated.
Theoretical studies of these phenomenon in degenerate BEC systems
\cite{moore,elena} pointed out the important role of correlations
and competitions among matter wave side modes, {\it i.e.} multi-mode
nature of even a single component condensate due to center of mass (CM)
motional effects. For a harmonically trapped condensate,
these multi-modes can be conveniently expressed
in terms of quantized motional states with equal energy spacing.
Theoretical investigations of light scattering from such
trapped atoms are complicated since both the elastic and
inelastic spectra can include contributions from many
different motional states.

In this article, by proposing a new identification of trapped
atomic side modes for light scattering from a plane wave excitation,
we attempt for a detailed interpretation of the recently
observed off-resonant super-radiant light scattering \cite{inouye}.
This constitutes an example of using light scattering
as a spectroscopic tool to probe properties of trapped
degenerate quantum gas. Our investigations show that quantum
statistics of the condensate can have a drastic effect on
the properties of scattered photons \cite{juha,you}. Our formalism
takes advantage of the recent success with atomic multi-modes
\cite{moore-zobay} to provide a clean interpretation of all aspects
of the experimental observations \cite{inouye}. Similar approaches
can also be used to clarify physical pictures of the
more recent BEC Bragg spectroscopy experiments \cite{stenger,kozuma}.

\section{formulation}
\begin{figure}[t]
\centerline{\epsfig{file=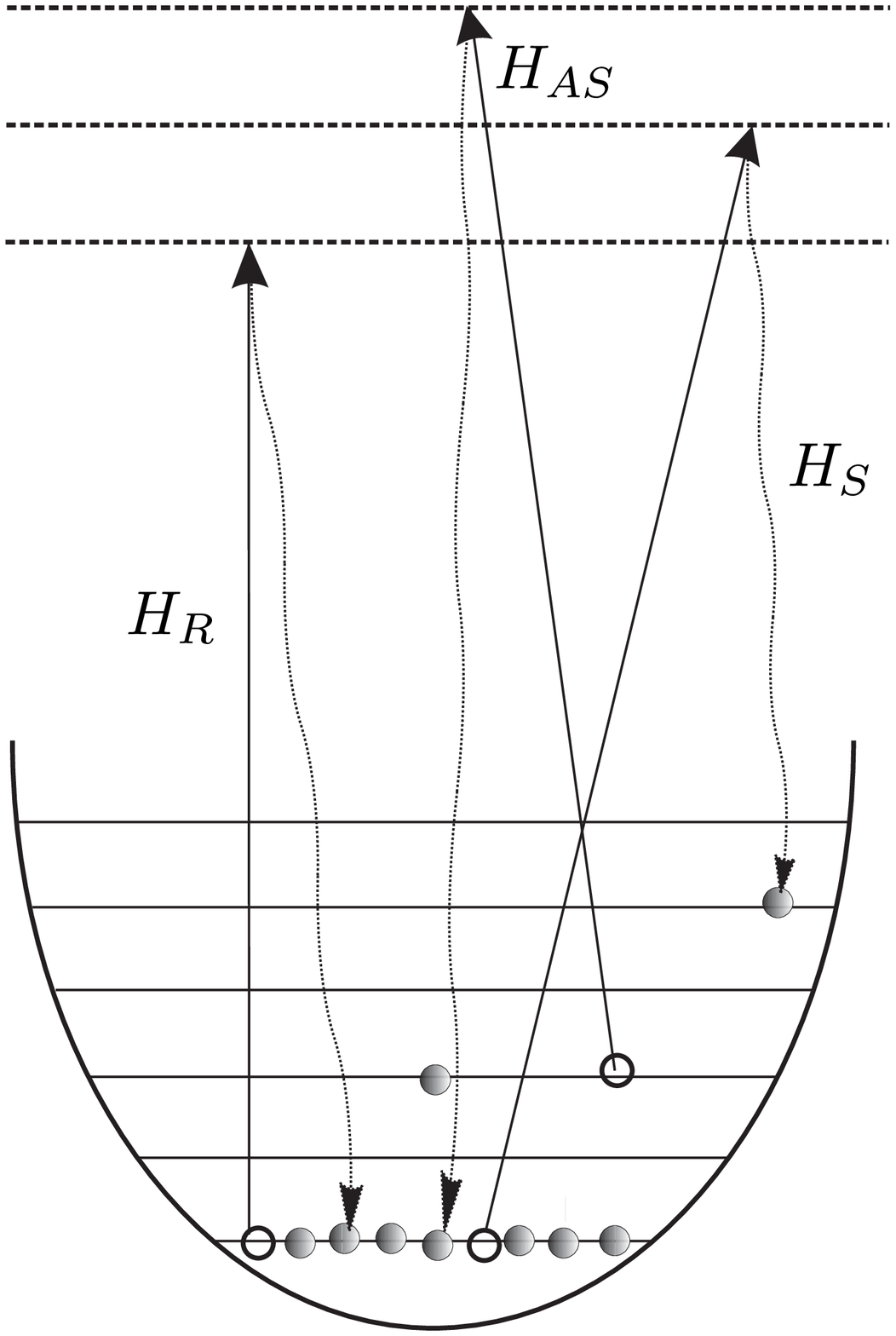,width=2.5in}\\[12pt]}
\caption{The geometry of the scattering arrangement
from trapped atoms. The filled ellipsoid represents the
atomic cloud with dimensions $L$ and $W$. Incident pump field
comes in along the short axis of the trap, while the geometry favors
the emission to be along the long axis of the trap.
Momentum conservation of the absorption and the subsequent emission
of photons results in the recoil of trapped atoms into
a wavepacket state parametrized by $\vec q$. }
\label{fig1}
\end{figure}

\begin{figure}[t]
\centerline{\epsfig{file=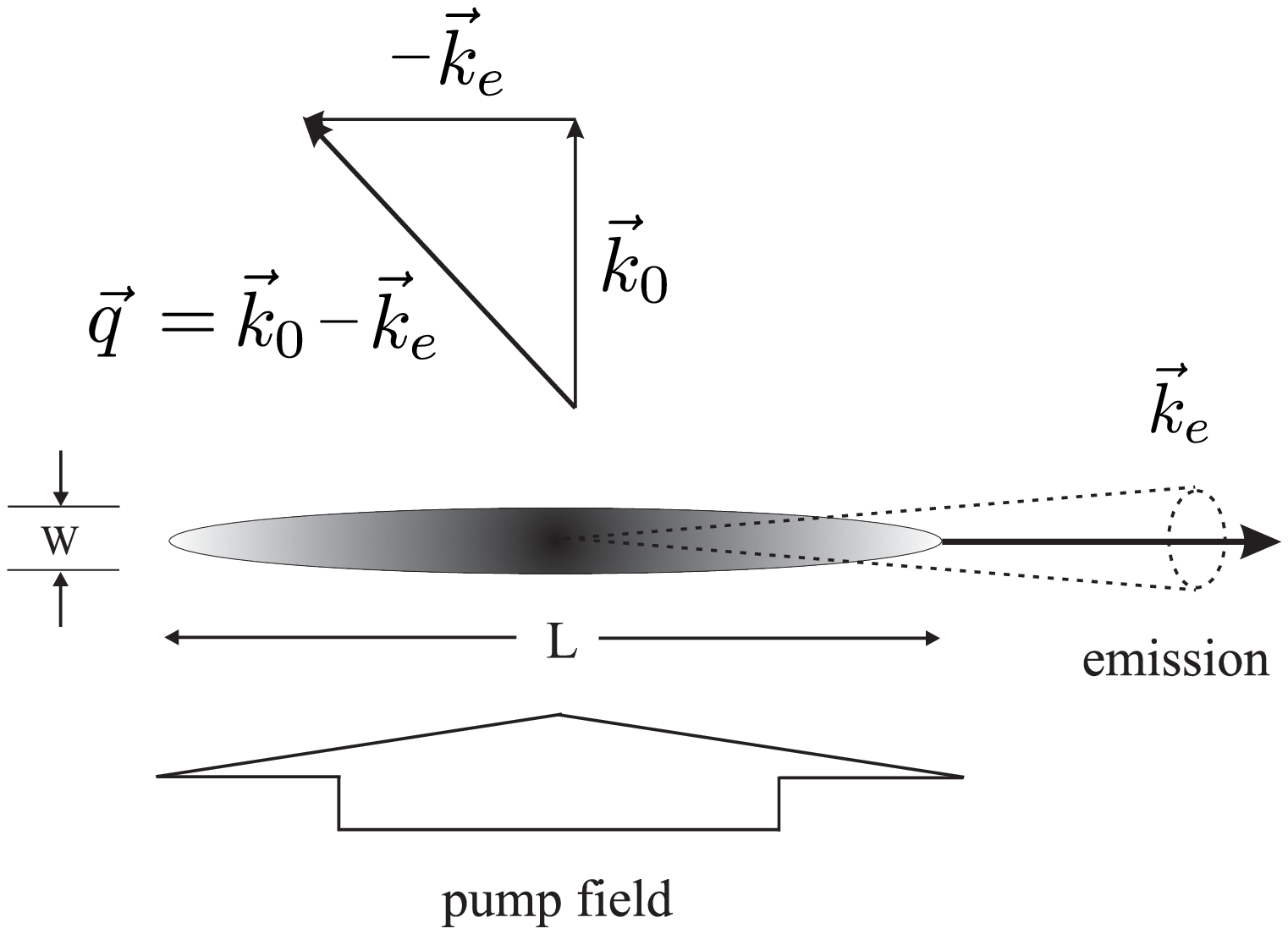,width=3.in}\\[12pt]}
\caption{The diagram for Rayleigh,
Raman Stokes and anti-Stokes scattering among
the motional states of trapped atoms. The solid lines
denote pump photons while dotted curve lines are
for scattered photons. Solid dots denote the presence of
an atom, and hollow dots denote the absence of atoms
due to scattering out of certain motional states.}
\label{fig2}
\end{figure}

Our model describes light scattering of trapped atoms
from excitation due to an intense far off-resonant plane wave pump field.
The proto-type system is illustrated in Fig. \ref{fig1}
as arranged in the MIT experiment \cite{inouye}.
The atomic sample is assumed dilute and atoms are of the alkali type
with a single valance electron. Two electronic states
[$g$ ($e$) for ground (excited)] are connected by a real
electronic dipole moment $d$. In units of $\hbar=1$ and
in length gauge, our model Hamiltonian takes the form
\begin{eqnarray}
H &=& \sum_{\sigma=g,e}\int d\vec r\,
\Psi^{\dag}_{\sigma}(\vec r)H_{A\sigma} \Psi_{\sigma}(\vec r) +
\int d\vec k \omega_{k}b_{\vec{k}}^{\dag}b_{\vec{k}} \nonumber\\
&+&\frac{1}{2}{\Omega_{0}(t)}\int d\vec r\,
\Psi_{e}^{\dag}(\vec r)e^{i\vec{k}_{0}\cdot\vec r}\Psi_{g}(\vec r)+h.c.\nonumber\\
&+&\int d\vec r \int d\vec k\, g(\vec{k})
\Psi_{e}^{\dag}(\vec r)e^{i\vec{k}\cdot\vec r}
b_{\vec{k}}\Psi_{g}(\vec r)+h.c. ,
\label{h1}
\end{eqnarray}
under dipole and rotating wave approximation.
$H_{A\sigma}=-\nabla^{2}/2M+V_{\sigma}(R)+E_{\sigma}$
is the atomic Hamiltonian consists of
CM kinetic energy (atomic mass $M$), trapping potential $V_{\sigma}$,
and electronic excitation energies in the rotating frame
$E_{g}=0, E_{e}=\Delta=\omega_{A}-\omega_0$.
$\Delta$ is large since the pump field at
frequency $\omega_0$ is far detuned
from the atomic transition frequency $\omega_{A}$.
The dipole interaction between an atom and the pump field
is described by a time dependent
Rabi frequency $\Omega_{0}(t)=\Omega_{0}{\cal T}(\gamma_{0}t)$,
with $\Omega_{0}$ the peak value and $\gamma_{0}$ the
temporal width of the envelope function ${\cal T}(\gamma_{0}t)$.
Both $\Omega_{0}$ and ${\cal T}(\gamma_{0}t)$ are chosen
to be real. The polarization and wave vector of the pump pulse
are denoted by $\vec{\epsilon}_{0}$ and $\vec{k}_{0}$, respectively.
The second term of Eq. (\ref{h1}) describes free
electromagnetic fields needed to consider inelastic photon
scattering. Their polarization index is
suppressed in the wave vector $\vec{k}$.
The degenerate atomic fields are described
by annihilation (creation) operators $\Psi(\vec r)$
[$\Psi^{\dag}(\vec r)$] satisfying appropriate commutators for
bosonic or classical Maxwell Boltzman statistics.
The operator $b_{\vec{k}}$ $(b^{\dag}_{\vec{k}})$ annihilates
(creates) a photon with wave vector $\vec{k}$,
polarization $\vec \epsilon_{\vec k}$, and energy
$\omega_{k}=ck-\omega_{0}$ (again in the frame rotating with
frequency $\omega_{0}$). Within the two state approximation,
dipole coupling between scattered field and an atom is
given by $g(\vec{k})=-i\sqrt{ckd^{2}/4\pi^{2}}\vec{\epsilon}_{0}^{\ast}
\cdot\vec{\epsilon}_{\vec{k}}$,
a slowly varying function of $k$. When the detuning
$\Delta$ is much larger than any other frequency scale in the
system, the excited state can be eliminated to obtain an
effective field theory within the ground atomic motional
state manifold \cite{moore},
\begin{eqnarray}
H&=&\int d\vec r\,
\Psi_{g}^{\dag}(\vec r)\left[H_{Ag}+\frac{|\Omega_{0}(t)|^{2}}
{2\Delta}\right]\Psi_{g}(\vec r)
+\int d\vec k\, \omega_{k}b_{\vec k}^{\dag}b_{\vec k}\nonumber\\
&+&\int d\vec r\int d\vec k\,g(\vec{k},t)
\Psi_{g}^{\dag}(\vec r)e^{i(\vec{k}-\vec{k}_{0})\cdot\vec r}
b_{\vec{k}}\Psi_{g}(\vec r)+h.c.\ .
\label{hpsi}
\end{eqnarray}
The AC stark shift term from all field modes except
the pump will be neglected in this study since we are interested
in regimes when the scattered field intensity remains small.
The effective coupling constant $g(\vec{k},t)=\Omega_{0}
{\cal T}(\gamma_{0}t)g(\vec{k})/2\Delta$ is now time dependent,
and it describes the scattering event in the ground motional state manifold,
{\it i.e.} the combined effect of absorbing a pump photon to the
excited state manifold followed by a spontaneous emission back to
the ground state again. In most cases $g(\vec{k},t)$ is a slowly
varying function of time. To simplify our discussion we consider a
temporal square shaped pulse and thus drop the
constant term ${|\Omega_{0}(t)|^{2}}/{2\Delta}$. In contrast to
strongly correlated many body systems the trapped atoms
are assumed non-interacting, allowing for
detailed investigations of their interaction with light fields.
The ground state atomic field operator can be expanded
in terms of trapped single atom wave function $\phi_{n}$ according to
$\Psi_{g}(\vec r)=\sum_{n}c_{n}\phi_{n}(\vec r)$. For a
harmonic trapping potential, $\phi_{n}(\vec r)$ is simply
the number basis states in position representation with
$n=(n_{x},n_{y},n_{z})$ a triplet index.
Atomic annihilation and creation operators
$c_{n}$ and $c_{n}^{\dag}$ obey bosonic algebra
$[c_n,c_m^\dag]=\delta_{nm}$ and $[c_n,c_m]=0$.
We then reduce the Hamiltonian (\ref{hpsi}) to
\begin{eqnarray}\label{hc3}
H&=&H_{0}+H_{R}+H_{AS}+H_{S}, \nonumber\\
H_{0}&=&\sum_{n}E_{n}c_{n}^{\dag}c_{n}
+\int d\vec k\,\omega_{k}b_{\vec{k}}^{\dag}b_{\vec{k}},\nonumber\\
H_{R}&=&\sum_{n}\int d\vec k\,g^{*}(\vec{k},t)
\eta_{n,n}(\vec{k}-\vec{k_{0}}) c^{\dag}_{n}b_{\vec{k}}^{\dag}c_{n}+h.c.,\nonumber\\
H_{AS}&=&\sum_{n}\sum_{m\in(E_m>E_n)}\int d\vec k\,g^{*}(\vec{k},t)
\eta_{n,m}(\vec{k}-\vec{k_{0}}) c^{\dag}_{n}b_{\vec{k}}^{\dag}c_{m}+h.c.,\nonumber\\
H_{S}&=&\sum_{n}\sum_{m\in(E_m<E_n)}\int d\vec k\,g^{*}(\vec{k},t)
\eta_{n,m}(\vec{k}-\vec{k_{0}}) c^{\dag}_{n}b_{\vec{k}}^{\dag}c_{m}+h.c. \ .
\end{eqnarray}
Figure \ref{fig2} is a pictorial display of characteristic
absorption and emission cycles for Rayleigh, Raman Stokes
and anti-Stokes processes.
The atomic energy is
$E_{n}=E_{0}+\omega_{n}$,
consisting of contributions from electronic ground state energy
$E_{0}=0$ and CM motion energy $\omega_{n}=\vec{\omega_{t}}\cdot\vec{n}$
with frequency $\vec{\omega_{t}}=(\omega_x,\omega_y,\omega_z)$
for a three dimensional trap.
The factor $\eta_{n,m}(\vec{K})=\langle n|
\exp{(-i\vec{K}\cdot\vec r)}| m\rangle$ represents
CM motional state dipole transition moment, and is
analogous to the Franck-Condon factor in a
di-atomic molecular transition. It is simply the matrix
element of displacement operator $D(\vec r)=\exp{(-i\vec{K}\cdot\vec r)}$
 in the number basis and depends on the total recoil momentum
$\vec{K}=\vec{k}_0-\vec{k}$ from the scattering cycle involving
absorption of a pump photon followed by an emission. Within the
ground motional states, it acts like a diffraction matrix since it
shifts atomic fields around in momentum space. To examine various
competing dynamical processes in light scattering described by Eq.
(\ref{hc3}) we separate the coupling term depending on the
energies of the two coupled (initial and final) motional states.
This leads to three types of scattering (as in Fig. \ref{fig2}):
1) the elastic Rayleigh scattering described by $H_R$ corresponds
to events within the same atomic motional states; 2) the Stokes
$H_S$ and 3) the anti-Stokes $H_{AS}$ Raman scattering into higher
(lower) energy motional states. We emphasise that Stokes and
anti-Stokes terms here corresponds to the same final electronic
state but with increased or decreased energy final motional
states. We may thus also call them as ``inelastic" Rayleigh
scattering.

Before a detailed discussion of the three type scattering events,
it is possible to get a crude picture of how each individual
interaction term contributes to the dynamics at
low temperatures and within a short time scale. When the
gas sample is at sufficient low temperatures
only low lying atomic motional states are densely occupied and
therefore their atomic fields can be approximated as
classical variables, while upper motional states are
sparsely populated and need to be treated quantum mechanically.
Within a short duration, an approximation involving
undepleted populations in lower motional states,
similar to the parametric pump approximation in nonlinear optical
multi-mode coupling, can be made.
We can then consider a single motional state (m) and a single
resonant scattering field mode with $\vec{k}=\vec{k}_{1}
(|\vec{k}_{1}|=k_{0})$, and assume all $N_{0}$
atoms were initially condensed in state $n=(0,0,0)$.
We then approximate $c_0\sim c_0^\dag\sim \sqrt{N_0}$, and
Hamiltonian (\ref{hc3}) is simplified to
\begin{eqnarray}
H&=&H_0+H_{R}+H_{AS}+H_{S}, \nonumber\\
H_0 &=& E_{m}c_{m}^{\dag}c_{m}+\omega_1b_{\vec{k}_1}^\dag b_{\vec{k}_1},\nonumber\\
H_{R}&=&g^{*}(\vec{k}_{1},t) \eta_{0,0}(\vec{k}_{1}-\vec{k_{0}})
N_{0}b_{\vec{k}_1}^{\dag}+O(c_m^\dag c_mb_{\vec{k}_1})+h.c. ,\nonumber\\
H_{AS}&=&g^{*}(\vec{k}_{1},t) \eta_{0,m}(\vec{k}_{1}-\vec{k_{0}})
\sqrt{N_0}b_{\vec{k}_1}^{\dag}c_{m}+h.c. ,\nonumber\\
H_{S}&=&g^{*}(\vec{k}_{1},t) \eta_{m,0}(\vec{k}_{1}-\vec{k_{0}})
\sqrt{N_0}c^{\dag}_{m}b_{\vec{k}_1}^{\dag}+h.c.\ .
\label{hc}
\end{eqnarray}
With this simplified model, $H_{R}$ becomes a displacement
operation for scattered field mode ($\vec k_1$); it describes the
generation of coherent photon fields in this mode.
For the general case of a thermally populated motional
state distribution, the total coherent
photon fields are distributed accordingly.
Terms of order $O(c_m^\dag c_mb_{\vec{k}_1})$ are
neglected since the second order quantum
processes described by $H_S$ and $H_{AS}$ are more important within
the short time period discussed here. We see that $H_{AS}$ resembles
an atom-polariton Hamiltonian, describing the generation
of atom-photon bound states; while $H_{S}$ takes the form of
a non-degenerate parametric amplifier Hamiltonian,
describing processes of gain, squeezing, and atom-photon entanglement.
It should be noted, however, even when both terms
$H_S$ and $H_{AS}$ are in existence, they can still be
grouped as a more general polariton Hamiltonian.
Thus, internal fields, {\it i.e.} scattered photons inside
the atomic sample can always be viewed as being part of
an atom-polariton system \cite{po}. Overall photon scattering and
emissions from such a system is complicated as
internal fields couple to external fields to cause
radiative decays of atom-photon polaritons.
Their periodic energy exchange implied by
$H_{AS}$ could exhibit Rabi type oscillations in the radiated fields.
If on the other hand when the motional state $m$
is initially unpopulated, oscillatory behaviors
will not appear since $H_{AS}$ can not
contribute during the early stages of dynamic interaction.

\begin{figure}[t]
\centerline{\epsfig{file=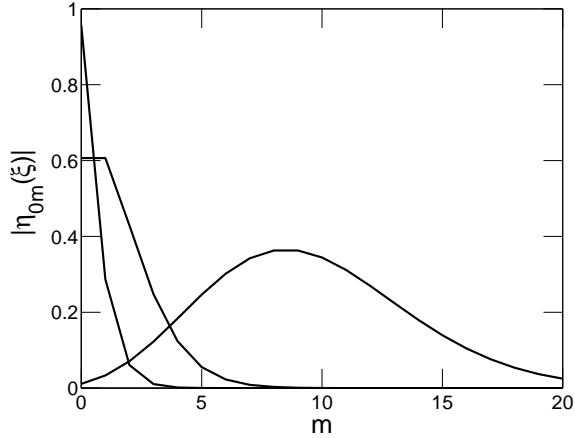,width=3.in}}
\caption{Typical values of the
Franck-Condon factor $|\eta_{0,n}(\xi)|$ for $\xi=0.3,1,$ and $3$.}
\label{fig3}
\end{figure}

In the following two sections, we will focus on situations
when $H_{S}$ is the dominant interaction term.
We start by outlining the required system conditions
to achieve this. First we observe that $H_{R}$
governs mostly small angle scattering while both Raman type
interactions give rise to mostly large angle scattering.
Mathematically this is due to the fact that diagonal elements
of the Franck-Condon factors are
sharply peaked around axis defined by pump wave vector
$\vec{k}_{0}$, while off-diagonal elements favor off-axial
scattering for traps of reasonable size $\sim$ a few times
the resonant wave length $\lambda_0$. The
interaction strengths of Rayleigh, Stokes, and anti-Stokes
processes all depend on the spatial distribution of atoms,
the initial ground state population $N_0$, and the
laser induced effective dipole interaction strength $g$.
In general these different factors compete and complicated
pictures of light scattering emerge. But for the
current model system, we find that the major
role is played by the geometry factor of the system
through $\eta_{n,m}$,
of which several off-diagonal elements are displayed in
Fig. \ref{fig3}.

\section{SMALL ANGLE SCATTERING AND OSCILLATORY SUPERRADIANCE}
Since $H_{R}$ is the leading order interaction term,
we will first consider its effect by discussing the
process of small angle scattering.
In this case, the system dynamics is determined
essentially by $H=H_{0}+H_{R}$. Using the property
$\eta_{n,n}(\vec{K})=\eta_{n,n}(-\vec{K})$, and
taking $g({\vec{k},t})=-i\gamma_{\vec{k}}$ with
$\gamma_{\vec{k}}$ real, we obtain the following
Heisenberg operator equations of motion
\begin{eqnarray}
\frac{d}{dt}\tilde{c}_{n}&=&\int d\vec k\, \gamma_{\vec{k}}\,
\eta_{nn}(\vec{K})P_{\vec{k}}\,\tilde{c}_{n}, \nonumber\\
\frac{d}{dt}P_{\vec{k}}&=&-i\omega_{k}Q_{\vec{k}}, \nonumber\\
\frac{d}{dt}Q_{\vec{k}}&=&-i\omega_{k}P_{\vec{k}}
-2\gamma_{\vec{k}}F(\vec{K}),
\label{eqr}
\end{eqnarray}
where we have defined auxiliary quadrature operators
$P_{\vec{k}}=b_{\vec{k}}-b_{\vec{k}}^{\dag}$,
$Q_{\vec{k}}=b_{\vec{k}}+b_{\vec{k}}^{\dag}$,
and the form function
operator $F(\vec{K})=\sum_{n}\eta_{n,n}(\vec{K})N_{n}$.
$\tilde{c}_{n}$ is a slowly variant form of ${c}_{n}$.
These equations and the auxiliary operators are similar to
those in the theory of collective atomic recoil laser (CARL) \cite{bonifacio}.
Since $P_{\vec{k}}^{\dag}=-P_{\vec{k}}$, we see that the
atom number operator $\tilde{c}_{n}^\dag\tilde{c}_{n}$
is time independent, a testimony of motional state number
conservation in Rayleigh scattering. The system of Eq.
(\ref{eqr}) is integrable and its solutions allow for
the calculation of the number of scattered photons
\begin{eqnarray}
\langle b_{\vec{k}}^{+}b_{\vec{k}}\rangle=2\gamma_{\vec{k}}^{2}
\left(\frac{\sin{\omega_{k}t/2}}{\omega_{k}}\right)^{2}
\left(\left|
\sum_{n}\eta_{nn}(\vec{K})N_{n}\right|^{2}+\sum_{n}\left|\eta_{nn}
(\vec{K})\right|^{2}\sigma^2(N_{n})\right).
\end{eqnarray}
We note that the amplitude of scattering light intensity depends on
atomic number fluctuations, the variance
$\sigma^2(N_n)=\langle N_n^2\rangle-\langle N_n\rangle^2$.
Depending on the gas temperatures, this
fluctuating part could dominate over the first coherent part
at larger scattering angles.
At very low temperatures, assuming a macroscopic condensed
atomic population in the lowest motional state $n=(0,0,0)$,
we find
\begin{eqnarray}
\langle b_{\vec{k}}^{\dag}b_{\vec{k}}\rangle=2\gamma_{\vec{k}}^{2}
\left(\frac{\sin{\omega_{k}t/2}}{\omega_{k}}\right)^{2}
\eta_{0,0}(\vec{K})^{2}[N_{0}^{2}+\xi (N_{0}+\epsilon
N_{0}^{2})],
\end{eqnarray}
where $\xi=0$ is for a classical gas, and
$\epsilon=0$ ($1$) stands for a coherent (Fock) state
of the Bose-Einstein condensate with an average of $N_0$ atoms.
The prefactor in the above two equations involves
$\sin(\omega_k t)$, a term from the time
dependent spectra of the square pump pulse.

At higher temperatures, it was previously calculated
that \cite{you},
\begin{eqnarray}
\sum_{n}\eta_{n,n}(\vec{K})N_{n}&\approx&
Ne^{-\frac{1}{2}K_{x}^2a_{x}^{2}/\beta\omega_{x}}
e^{-\frac{1}{2}K_{y}^2a_{y}^{2}/\beta\omega_{y}}
e^{-\frac{1}{2}K_{z}^2a_{z}^{2}/\beta\omega_{z}},\\
\sum_{n}\left|\eta_{n,n} (\vec{K})\right|^{2}
\sigma^2(N_{n})&\approx& N^{2}\left(
\frac{\beta^{3}\omega_{x}\omega_{y}\omega_{z}}{8}\right)
e^{-\frac{1}{2}K_{x}^2a_{x}^{2}\beta\omega_{x}}
e^{-\frac{1}{2}K_{y}^2a_{y}^{2}\beta\omega_{y}}
e^{-\frac{1}{2}K_{z}^2a_{z}^{2}\beta\omega_{z}},
\end{eqnarray}
where $\beta=1/k_BT$ is proportional to the inverse temperature
and the trap ground state size is
$a_{\alpha}=\sqrt{(1/2M\omega_{\alpha})}$ in direction $\alpha=x,y,z$.
We can readily see that the incoherent part contributes mostly
at higher temperatures while the
coherent part is more effective at lower temperatures. In order to
suppress incoherent large angle scattering we need to satisfy
$\frac{1}{2}K_{\alpha}^2a_{\alpha}^{2}\beta\omega_{\alpha}\gg1$,
{\it i.e.} $E_{\alpha}\beta/2\gg1$, where $E_{\alpha}=K_{\alpha}^2/2M$
is the one dimensional recoil energy. We emphasize that
this result is independent of the shape of trapped
atomic sample. In terms of recoil temperature $T_R=K^2/2Mk_B$,
a sufficient condition is $T\ll0.5T_{R}$.
At such low temperatures coherent Rayleigh scattering dominates.
In order to suppress coherent scattering, we
can now take advantage of trap size parameters.
For a cylindrical sample with $L$ ($W$) the long (short)
axis length as in Fig. \ref{fig1}, the
coherent scattering is controlled by
\begin{eqnarray}
|\eta_{0,0}(\vec{k}-\vec{k}_{0})|^2=e^{-k^{2}L^{2}\sin^{2}{\theta}\cos^{2}{\phi}}
e^{-k^{2}W^{2}\sin^{2}{\theta}\sin^{2}{\phi}}
e^{-W^{2}(k\cos{\theta}-k_{0})^{2}},
\label{f00}
\end{eqnarray}
where $z$ axis is chosen to be along the pump field
direction and $\theta,\phi$ are polar and azmithual
angles of the scattering direction. Typical geometries
of current traps are $L=W\sim 10$ ($\mu$m) for spherical traps
and $L=10W\sim 200$ ($\mu$m) for cigar shaped traps.
Putting these into Eq. (\ref{f00}), we see that the
cigar shaped geometry with $L\gg W$ is more effective in
suppressing the overall coherent Rayleigh scattering.

Concluding this section we note that under optimal
conditions, it is possible to significantly suppress
the coherent Rayleigh scattering. Therefore effects of the
other wise higher order Raman processes can be made dominant.
As reasoned before we can also ignore contributions
from anti-Stokes terms at sufficient low temperatures and
for short pulse excitation. Thus we shall first develop a simple
model considering $H_{S}$ as the only dominant
mechanism in describing the directional and exponential superradiance.
A more complete treatment including anti-Stokes
processes will be considered in
Sec. V using a generalized atomic side mode formalism.

\section{LARGE ANGLE SCATTERING AND EXPONENTIAL GAIN BEHAVIOR}
In this section we consider large angle Raman
scattering processes when small angle Rayleigh scattering
is suppressed. In the low temperature limit
when all atoms are condensed into the ground motional state,
we can neglect the ground state atom number fluctuations
and approximate $c_{0}=\sqrt{N_{0}}$ as a classical field.
The subsequent system dynamics for light scattering can
be described by the Hamiltonian
\begin{eqnarray}
H&=&\sum_{n>0}E_{n}c_{n}^{\dagger}c_{n}+\int d\vec k\,
\omega_{k}b_{\vec{k}}^{\dagger}{b_{\vec{k}}} \nonumber\\
&+&\sqrt{N_{0}}\int d\vec k \left[g^{*}(\vec{k},t)
\sum_{n>0}\eta_{n,0}(\vec{k}-\vec{k}_{0})
c_{n}^{\dagger}b_{\vec{k}}^{\dag}+h.c. \right].
\end{eqnarray}
In this limit, further insight into this
problem can be obtained by introducing atomic side mode operators
\begin{eqnarray}
f(\vec{q})=\sum_{n>0}\eta_{0,n}(\vec{q})c_{n}.
\end{eqnarray}
Physically these are wave packet operators in the ground
state manifold due to absorption of a pump photon followed
by emitting a spontaneous photon, resulting in a net
momentum transfer of $\vec q$. Because momentum conservation
is in general violated during photon absorption and emission
among two distinct initial and final motional states.
These wavepacket operators are the best compromise one can construct
to reflect the momentum conservation law for
starting in the ground motional state before the absorption and
emission cycle. Mathematically, they satisfy the
following commutation relation,
\begin{eqnarray}
[ f(\vec{q}),f^{\dag}(\vec{p})]
=\eta_{00}(\vec{q}-\vec{p})
-\eta_{00}(\vec{q})\eta_{00}(-\vec{p})
\equiv {\cal D}_{\vec{q}}(\vec{p}).
\end{eqnarray}
Of particular interest is the special case of
$[f(\vec{q}),f^{\dagger}(\vec{q})]=1-\eta_{00}^2(\vec{q})$.
These operators can thus be visualized as a deformation on
the Weyl-Heisenberg algebra [H(4)] of the original
$c_{n}$ operators. On the other hand, if we were to
keep $c_{0}$ as an operator, we would
need to consider $X_{n0}=c_{n}^{\dagger}c_{0}$ operators \cite{gardiner}.
We find that $c_{n}$ can be realized as a bosonic
representation of transition operators $X_{n0}$, which obeys
a U(3) Lie algebra. A similar deformation on the U(3) could
also be proposed. More generally, a class of side mode
operators from an arbitrary motional state, not limited to $n=0$
could also be considered as in Sec. V. They represent collective
recoil modes corresponding to various transitions from
any motional state to higher or lower motional states
with a net recoil momentum $\vec{q}$ during the photon absorption
and emission cycle.

Neglecting the Doppler effect but keeping the
recoil energy term $\omega_R(\vec K)=K^2/2M$, we can show that
\begin{eqnarray}
\sum_{n>0}E_{n}\eta_{0,n}(\vec{q})c_{n}\approx
[E_0+\omega_R(\vec{q})]f(\vec{q}).
\end{eqnarray}
The equations of motion for the operators are then given by
\begin{eqnarray}
i\frac{d}{dt}f(\vec{q})&=&\omega_R(\vec{q})f(\vec{q})+\sqrt{N_{0}}
\int d\vec{k}\,
g^{*}(\vec{k},t)b_{\vec{k}}^{\dagger}\,
{\cal D}_{\vec{q}}(\vec{k}_{0}-\vec{k}),\label{fdot}\\
i\frac{d}{dt}b_{\vec{k}}&=&\omega_{k}b_{\vec{k}}+\sqrt{N_0}\,g^{*}(\vec{k},t)
f^{\dagger}(\vec{k}_{0}-\vec{k}).
\label{bdot}
\end{eqnarray}
We can proceed with standard technique to
eliminate the scattered field modes by substituting the
formal solution of Eq. (\ref{bdot}) into Eq. (\ref{fdot}).
This yields
\begin{eqnarray}
\frac{d}{dt}f(\vec{q})+i\omega_R(\vec{q})f(\vec{q})
=F_{L}^{\dag}+N_{0}\int d\vec k\,
g^{*}(\vec{k},t){\cal D}_{\vec{q}}(\vec{k}_{0}-\vec{k})\int_{0}^{t}d\tau
f(\vec{k}_{0}-\vec{k},t-\tau)g(\vec{k},t-\tau)e^{i\omega_{k}\tau},
\label{fsmall}
\end{eqnarray}
where the Langevin noise operator $F_{L}(t)$, representing the
effect of vacuum fluctuations through the initial scattered field
operators $b_{\vec{k}}(0)$, is introduced as in \cite{moore}
\begin{eqnarray}
F_{L}(t)=i\sqrt{N_{0}}\int d\vec k\,
g(\vec{k},t){\cal D}_{\vec{q}}(\vec{k}_{0}-\vec{k})b_{\vec{k}}(0)e^{-i\omega_{k}t}.
\label{fl}
\end{eqnarray}
This noise term is responsible for triggering the
super-radiant emission from the gas sample. It is also needed
to satisfy thermal dynamic requirement of fluctuation
and dissipation theorem. Since its magnitude is of
a lower order in $\sqrt{N_0}$, it can be neglected in a
semi-classical description of the growth behavior
of a small signal gain as in Ref. \cite{moore}.
In the Markov approximation and with typically small recoils
[$\omega_R(\vec{q})\rightarrow0$], we can ignore the slow
time dependence of $g(\vec{k},t)$ to obtain
\begin{eqnarray}
\frac{d}{dt}f(\vec{q})=\pi N_{0}\int d\vec k\,
|g(\vec{k},0)|^{2}{\cal D}_{\vec{q}}(\vec{k}_{0}-\vec{k})
f(\vec{k_{0}}-\vec{k})\delta(\omega_k).
\label{fav}
\end{eqnarray}
This expression can be further simplified by noting that
$D_{\vec{q}}(\vec{k}_{0}-\vec{k})$ limits the scattering
to directions around the end firing modes
$\vec{k_{0}}-\vec{q}$ as illustrated
in Fig. \ref{fig1}. We then obtain a simple
exponential gain behavior for atomic
side mode operator according to
$f(\vec{q},t)=\exp{(N_{0}G_{\vec{q}}\,t/2)}f(\vec{q},0)$
with the gain parameter given by
\begin{eqnarray}
G_{\vec{q}}=\pi\int d\vec k\,
|g(\vec{k},0)|^{2}{\cal D}_{\vec{q}}(\vec{k}_{0}-\vec{k})\delta(\omega_k).
\end{eqnarray}
It is interesting to note that this result
is essentially the same as obtained in Ref. \cite{moore}
except for the difference in ${\cal D}_{\vec{q}}(\vec p)$.
In fact, the condensate shape function was defined
in Ref. \cite{moore} as $\rho_{\vec{q}}(\vec{k})=\int
d\vec r\,|\phi_0(\vec r)|^{2}\exp[-i(\vec{k}-\vec{k}_0+\vec{q})\cdot\vec r]$.
It is related our function defined here by
${\cal D}_{\vec{q}}(\vec{k}_0-\vec{k})=\rho_{\vec{q}}(\vec{k})
-\rho_{\vec{q}}(\vec{k}_0)\rho_{0}(\vec{k})$.

A more rigorous treatment would include explicitly
the role of atomic quantum statistics.
Treating $N_{0}$ as an operator, and defining the number
operators of recoiled atoms as
$n_{\vec{q}}(t)=\left<f^{\dagger}(\vec{q},t)f(\vec{q},t)\right>$,
we find
$n_{\vec{q}}(t)={\cal G}_{\vec{q}}(t)n_{\vec{q}}(0)$,
where the growth function is now defined to be
${\cal G}_{\vec{q}}(t)=\left<\exp{(G_{\vec{q}}tN_{0})}\right>$.
In normally ordered form it can be written as
\begin{eqnarray}
{\cal G}_{\vec{q}}(t)=\langle e^{tG_{\vec{q}}N_{0}}\rangle
=\sum_{j=0}^{\infty}[e^{tG_{\vec q}}-1]^{j}
\langle c_{0}^{\dag j}c_{0}^{j}\rangle/{j!}.
\end{eqnarray}
Clearly, for different quantum statistical distributions,
this implies different growth behavior.
If we choose the initial condensate to be a coherent
state in the motional ground state with
an amplitude $\alpha$
such that $\langle c_{0}^{\dag}c_{0}\rangle=|\alpha|^2=\langle{N}_{0}\rangle$,
we obtain
\begin{eqnarray}
{\cal G}_{\vec{q}}(t)=\exp[(e^{G_{\vec{q}}t}-1)\langle{N}_{0}\rangle].
\end{eqnarray}
While for a Fock state distribution, the same result as in the
semi-classical case applies. We note that at early times ($t\ll
1/G_{\vec{q}}$) both growth curves give identical results
irrespective of the atom number statistics of the condensate. At
longer times, however, an initial condensate in a coherent state
causes side modes to grow faster than an initial Fock state
condensate. In Fig. \ref{fig6}, we compare the different growth
curves for both cases. Given the same number of condensate atoms,
the corresponding super-radiant pulse is then shorter for a
coherent state. For the recent MIT experiment \cite{inouye}, it
was estimated that with a laser intensity of $10-100$ (mW/cm$^2$),
$G_{\vec{q}}\le 10^{-4}$. Thus for all experimental observed
duration, the growth curve is the same irrespective of atom number
statistics. We note that if the MIT experiment were operated with
a higher pump laser intensity, condensate atom number fluctuations
could be probed. In contrast to the small angle Rayleigh
scattering where number fluctuations appear as amplitude
fluctuations, the large angle Raman scattering studied in this
section carries information directly related to condensate number
fluctuations.

\section{SEQUENTIAL SUPERRADIANCE}
In the previous section we considered short pulse Stokes Raman
scattering in which processes starting from the motional ground
state $n=(0,0,0)$ dominate. As a result, the momentum distribution
of atoms, sharply peaked around the center-mode $\vec{p}=\vec{0}$ initially,
is modified by the appearance of side mode
peaks around $\vec{p}=\vec{k}_{0}\pm\vec{k}_{e}$, where
$\pm\vec{k}_{e}$ denote the two end firing modes, {i.e.}
emissions along the two ends of an elongated condensate as in Fig. \ref{fig1}.
This situation is reminiscent of earlier studies of
superfluorescence from an extended and inverted medium
\cite{polder,skriban,bonifacio2,gibbs,bowden1,bowden2,you2}.
When the Fresnel number of the system, defined as
${\cal F}=\pi (W/2)^2/\lambda_0 L$, is of order unity,
a description of the emission can be made in terms
of the two end firing modes \cite{polder}.
For the MIT experiment with $W\sim$ 20 ($\mu$m),
$L\sim$ 200 ($\mu$m), and $\lambda_0\sim$ 0.6 ($\mu$m),
the condition ${\cal F}\sim 1$ is indeed satisfied.
With the pump incident along the narrower side of the
condensate, possibilities exist for mode couplings into
similar end firing modes. This in turn causes recoiling
atoms to couple with side modes of $\vec{p}=\vec{k}_{0}\pm\vec{k}_{e}$,
and even higher order side modes if the pump stays on for
a long period of time.
Since $\eta_{0,m}(\vec{k}_{0}\pm\vec{k}_{e})$ is peaked around
certain motional state $m_1$, this will be reflected in the
measured density profile as effective couplings
populate state $m_1$ from condensate atoms in state $m=(0,0,0)$.
With the wave packet formulation, a breadth of ground motional
states centered around $m_1$ is affected, and it is termed
collectively as a side mode to the original condensate.
This physical picture is further illustrated by the side mode
lattice as given in Fig. \ref{fig5}, where the most important
coupling terms from the Hamiltonian Eq. (\ref{hc}) have been
selected. We can then truncate the number of nodes involved in
the lattice depending on the duration of the pulse
to study light scattering dynamics. We introduce generalized
side mode operators as follows,
\begin{eqnarray}\label{fn}
f_{n}(\vec{q})=\sum_{m}\eta_{n,m}(\vec{q}-\vec{p})f_{m}(\vec{p}),
\end{eqnarray}
with $f_{n}(0)=c_{n}$. Their commutator is evaluated to be
\begin{eqnarray}\label{algebra}
[f_{n}(\vec{q}),f_{m}^{\dag}(\vec{p})]=\eta_{n,m}(\vec{q}-\vec{p}).
\label{com1}
\end{eqnarray}
We can then transform the system Hamiltonian
into these side modes to obtain,
\begin{eqnarray}
H&=&\sum_{n}[E_{n}+\omega_R({\vec q})]f_{n}^{\dag}(\vec{q})f_{n}(\vec{q})\nonumber\\
&+&\sum_{n,m}\vec v_{R}\cdot\vec{P}_{n,m}
f_{n}^{\dag}(\vec{q})f_{m}(\vec{q}) + \int
d\vec k\,\omega_{k}b_{\vec k}^{\dag}b_{\vec k} \nonumber \\ \
&+&\sum_{n} \int d\vec k\, [ g^{*}(\vec{k},t)
f^{\dag}_{n}(\vec{q}-\vec{k}+\vec{k}_{0})b_{\vec{k}}^{\dag}f_{n}(q)+h.c. ].
\end{eqnarray}
In the second term, $\vec{v}_{R}=\vec{q}/M$ is the recoil
velocity and $\vec{P}_{n,m}$ is the CM momentum matrix
elements between the Fock basis $|n\rangle$ and $|m\rangle$.
This term couples $f_{n}$ and $f_{n+1}$ and in effect
describes the hopping between nearest neighbor motional states.
In our discussion below we will neglect this nearest neighbor
motional coupling as it is small in the short time scale of
the MIT experiment \cite{inouye,moore-zobay}. Keeping only the
two end firing photon modes around $\pm\vec{k}_{e}$,
we can use Eq. (\ref{fn}) to express side modes
around them according to
\begin{eqnarray}
f_{n}(\vec{q}+\vec{k}_{0}-\vec{k})=\sum_{m}\eta_{n,m}(\mp\vec{k}_{e}-\vec{k})
f_{m}(\vec{q}+\vec{k}_{0}\pm\vec{k}_{e}).
\end{eqnarray}
This implies that by simply examining the dynamics of
the two end firing side modes, we can also gain valuable
understanding of the behavior of other side modes around them.
We further simplify the problem by taking only the
diagonal Franck-Condon factors, justified by the reasonably
small Fresnel number ${\cal F}$. This allows us to use
$f_{n}(\vec{q}+\vec{k}_{0}-\vec{k})\approx
\eta_{n,n}(\mp\vec{k}_{e}-\vec{k})
f_{n}(\vec{q}+\vec{k}_{0}\pm\vec{k}_{e})$ as
$\eta_{n,m}(\mp\vec{k}_{e}-\vec{k})\sim \delta_{nm}$
for modes near the two end firing ones with $\vec{k}\approx \mp\vec{k}_{e}$.
Since only $f_{0}(0)$ is initially occupied, this Hamiltonian
then couples side
modes with $f_{0}(\vec{q})$ for $\vec q=0,\vec{k}_{0}\pm\vec{k}_{e},
2\vec{k}_{0},2(\vec{k}_{0}\pm\vec{k}_{e})$,$\cdots$ etc. through an
infinite hierarchy of equations of nearest neighbor coupling on the
triangular lattice as in Fig. \ref{fig5}.

We now discuss effects of the second order side modes.
Since the central side mode at
$2\vec{k}_{0}$ is coupled to two first order side modes at
$\vec{k}_{0}\pm\vec{k}_{e}$, it will grow faster than
other second order ones at
$2(\vec{k}_{0}\pm\vec{k}_{e})$. Therefore, as indicated in
Fig. \ref{fig5}, we close the system of coupled nodes by
considering the 4 lattice nodes at $0,\vec{k}_{0}\pm\vec{k}_{e},2\vec{k}_{0}$,
which are connected with solid lines.
Pulses with longer duration, however, would result in
populations grow at higher order lattice nodes.
After free expansion on turning off the trapping potential,
this particular lattice structure is in fact directly
observed in the MIT experiment \cite{inouye}.
The effective Hamiltonian is now given by
\begin{eqnarray}
H &&= \sum_{\vec q}\omega_R({\vec q})f_0^\dag(\vec q)f_0(\vec q)+
\int d\vec k\,\omega_{k}b_{\vec k}^{\dag}b_{\vec k}\nonumber\\
&&+\sum_{\vec{q},\epsilon=\pm}
 [ g^{*}(\epsilon\vec{k}_{e},t)B_{\epsilon}^{\dag}
f_0^{\dag}(\vec{q}-\epsilon\vec{k}_{e}+\vec{k}_{0})f_0(\vec q)+h.c.],
\label{eqcc}
\end{eqnarray}
where $\vec{q}$ runs over lattice sites
$0,\vec{k}_{0}\pm\vec{k}_{e}$ such that we have a truncated
problem on the first diamond. The lattice only runs in one
direction with a positive $\vec k_0$ because of the plane wave
excitation from one side. The first term of Eq. (\ref{eqcc}) is
due to the recoil shift, and can be eliminated by transforming to
an interaction picture. $g(\vec{k},t)$ is a slowly varying
function of $\vec{k}$ and is replaced with its value at $\vec
k=\vec k_e$ and taken out of the integration over $\vec{k}$. The
emission photon wave packet operator is defined as
$B_{\epsilon}=\int d\vec
k\,\eta_{0,0}(\vec{k}-\epsilon\vec{k}_{e})b_{\vec{k}}$ for $\vec
k$ near $\pm\vec k_e$. The evolution of its corresponding
intensity naturally gives the photon scattering distribution
averaged over repeated single-shot experiments. As a collective
field operator, $B_{\epsilon}$ takes into account the multi mode
but directed (end firing) nature of the scattered field
\cite{bonifacio}. Although mathematically one obtains
\begin{eqnarray}
[B_\epsilon,B_{\epsilon'}^\dag]
&&=\delta_{\epsilon\epsilon'} {\pi^{3/2}\over a_x a_y a_z}.
\end{eqnarray}
We take $[B_\epsilon,B_{\epsilon'}^\dag]=\delta_{\epsilon\epsilon'}$
since the choice of keeping only end firing modes
constrains $\vec k$ to be around $\pm \vec k_e$. Thus one simply
has $B_{\epsilon}\sim b_{\epsilon\vec k_e}$. Other modes around
the end firing ones in $B_\epsilon$ only contributes to a
renormalization of the coupling constant $g$ which we take to
be phenomenological.
Terms involving $f[2(\vec{k}_{0}\pm\vec{k}_{e})]$ are ignored
in this study based on arguments of short pump pulses
and low atom number populations.

It is now useful to introduce a more concise notation
$a_0, a_{\pm}$, and $a_2$ for
$f_0(0),f_0(\vec{k}_{0}\pm\vec{k}_{e})$, and $f_0(2\vec{k}_{0})$.
We also treat these operators as commuting with
each other as an approximation to Eq. (\ref{com1})
in the limit of large $|\vec q-\vec p|$. Their Heisenberg
operator equations can be derived. It turns out their
dynamics is more transparently expressed in terms of
the population operators $I_{\pm}=B_{\pm}^{\dag}B_{\pm},
N_{\epsilon}=a_{\epsilon}^{\dag}a_{\epsilon}$, and
coherence operators $R_{\epsilon\epsilon'}=a_{\epsilon}^{\dag}a_{\epsilon'}$.
We find
\begin{eqnarray}
\frac{d}{dt}I_\pm&=&igR_{0\mp}B_{\pm} + igR_{\pm2}B_{\pm}+h.c., \nonumber\\
\frac{d}{dt}N_0&=&-igR_{0-}B_{+} - igR_{0+}B_{-}+h.c., \nonumber\\
\frac{d}{dt}N_\pm&=&igR_{0\pm}B_{\mp} - igR_{\pm2}B_{\pm}+h.c., \nonumber\\
\frac{d}{dt}N_2&=&igR_{+2}B_{+} + igR_{-2}B_{-}+h.c. \ .
\label{eqc}
\end{eqnarray}
In deriving this and other equations to follow,
we have consistently used an operator ordering with
atomic operators always to the left of all photon
operators. A careful analysis Eq. (\ref{eqc}) reveals
that the following two conservation laws are observed
\begin{eqnarray}
N_0(t)+N_-(t)+N_+(t)+N_2(t)&=&{\cal C}_1,\nonumber\\
N_0(t)-N_2(t)+I_+(t)+I_-(t)&=&{\cal C}_2,
\end{eqnarray}
with constants ${\cal C}_{j}$ determined by initial conditions.
In fact, the second
conservation law immediately implies possibility of
sequential superradiance. In the early stages
of the applied pump when condensate depletion is small,
scattered light intensity remains low, although
gradually increasing.
Eventually, the rapid decay of the condensate populations
($N_0\to N_{\pm}$)
sets in and the total light intensity starts to
increase sharply. The scattering losses and
absorption then causes the light intensity to decay
and finally vanish (when $N_0$ empties) while
$N_2$ remains small. The dynamics upto this point
is indeed equivalent to a system without the
presence of $N_2$ term, and is simply a parametric
amplification process. On the other hand, for long pulses with
sufficient intensity, the now populated $N_{\pm}$ nodes
start to dynamically populate node $N_2$. Thus allowing
for an revival of the scattered light intensity.

In the following, we shall be most interested
only in the population dynamics. Instead of solving the full set
of Eq. (\ref{eqc}), we assume equal population distribution
among the symmetric nodes of Fig. \ref{fig5},
{\it i.e.} treating nodes of $\vec k_0\pm\vec{k}_e$ as
equivalent. We can then define
$I=I_{+}+I_{-}$ and $N_1=N_{+}+N_{-}$.
This allows for the consideration of an effective set
of equations
\begin{eqnarray}
\frac{d}{dt}I &=& igR_{01}B + igR_{12}B +h.c., \nonumber\\
\frac{d}{dt}N_0&=&-igR_{01}B + h.c.,\nonumber \\
\frac{d}{dt}N_1&=&igR_{01}B-igR_{12}B+h.c.,\nonumber \\
\frac{d}{dt}N_2&=&igR_{12}B+h.c.,
\label{eqp}
\end{eqnarray}
with $B$ now denoting either of $B_{\pm}$.
The same conservation laws $N_0+N_1+N_2={\cal C}_1$ and
$N_0-N_2+I={\cal C}_2$ apply.
We can proceed to eliminate the scattered field
operator $B$ from the population dynamics Eq. (\ref{eqp})
by using the standard technique of substituting in
the formal solution for $b_{\vec k}(t)$. A Markovian
version of closed equations will be obtained this way
later which allows for a direct numerical simulation
in terms of averaged variables \cite{moore}.
Alternatively, we choose to develop a hierarchy of
equations for various operator moments first.
It is illuminating to follow both methods
and compare their results in the end.

We now introduce operators $X=R_{01}B$, $Y=R_{12}B$,
$Z=R_{12}R_{10}$, and $W=R_{02}B^{2}$. This is needed for a more
rigorous treatment of operator correlations, a procedure similar
to the random phase approximation \cite{rpa}. A trivial first
order decorrelation approximation between matter and field
$\left<R_{01}B\right>= \left<R_{01}\right>\left<B\right>$ would
have neglected too much correlation. By forming products involving
at least 4 operators from $b_{\vec k}^\dag$, $b_{\vec k}$,
$a_\epsilon^\dag$, and $a_\epsilon$, and making corresponding
decorrelation approximations
$\left<IX\right>=\left<I\right>\left<X\right>$, etc, we aim for a
closed set of equations. Although systematic, this approximation
procedure is not necessarily self-consistent as neglected higher
order correlated terms may be of the same magnitude of the kept
moments. There is solid evidence that this is a good approximation
for super-radiant systems \cite{new}. Our aim is to obtain a
truncated set of equations involving only limited number of higher
order operator moments which are relevant to the population
dynamics. We start by taking the averages of Eq. (\ref{eqp}), the
right hand side then immediately motivates the introduction of
operators $X,Y,Z$, and $W$. Upon averaging over their dynamic
operator equations, even higher order moments in general appear.
We then follow the decorrelation approximation as outlined above,
and only keep factorized products of already introduced operator
moments. More details of this higher order decorrelation
approximation can be found in previous treatments of
superfluorescence \cite{new}. Finally we drop the sign
$\langle.\rangle$ for averages and replace $g$ by $-i\gamma$ with
$\gamma$ real to obtain
\begin{eqnarray}
\frac{d}{dt}I&=&-\kappa I+\gamma(X+Y+c.c.),\nonumber\\
\frac{d}{dt}N_0&=&-\gamma(X+c.c.),\nonumber\\
\frac{d}{dt}N_1&=&-{\cal L}
N_1+\gamma(X-Y+c.c.),\nonumber\\
\frac{d}{dt}N_2&=&\gamma(Y+c.c.),
\label{nm1}
\end{eqnarray}
equations for the higher order operator moments
\begin{eqnarray}
\frac{d}{dt}X&=&-\gamma_\bot X
+\gamma[N_0(1+N_1)+I(N_0-N_1)+Z^*-W],\nonumber\\
\frac{d}{dt}Y&=&-\gamma_\bot
Y+\gamma[N_1(1+N_2)+I(N_1-N_2)+Z+W],\nonumber\\
\frac{d}{dt}Z&=&-\gamma_\bot
Z-2\gamma[X^*(N_2-N_1)+Y(N_1-N_0)],\nonumber\\
\frac{d}{dt}W&=&-\gamma_\bot W-2\gamma[I(Y-X)-N_0Y-X(1+N_2)],
\label{nm2}
\end{eqnarray}
as well as equations for their complex conjugates.

This system of twelve equations can be compared with the
Maxwell-Bloch equations describing superfluorescence from a sample
of coherently pumped three level atoms \cite{bowden1,bowden2}.
Similar to current trapped BEC systems, earlier superfluorescence
studies also assumed cigar (or pencil) shaped gas distribution,
albeit with much larger volumes. In those earlier studies, the
pump pulse is typically along the long axis direction of the gas
sample that is typically about several centimeters long
\cite{skriban,gibbs}. Theoretical analysis included both
propagation retardation and transverse diffractive effects
\cite{watson,mattar}. In the recent MIT experiment, the sample
size is much smaller and the pump pulse is along the transverse
direction of the long axis \cite{inouye}. Nevertheless, at higher
pump power super-radiant pulse shapes from the BEC display
multiple pulses or ringing effects similar to earlier hot gases
experiments \cite{mandel}. In this respect, sequential
superradiance may arguably be considered as a temporal analog of
spatial effects observed earlier although the mechanism is clearly
different. More recent experiments observed temporal ringing as an
intrinsic property in hot gas superfluorescence \cite{heinzen}.
Here in the BEC superradiance system ringing can also be
understood in terms of the cascading structure on the lattice
(Fig. \ref{fig5}) as opposed to among different electronic levels
\cite{kumar1}.

In the above Eqs. (\ref{nm1}) and (\ref{nm2}), we have introduced
phenomenological parameters $\gamma_\bot$,
${\cal L}$, and $\kappa$ which are respectively atomic side mode
dephasing rate, decay rate of first atomic side modes
due to coupling with excluded nodes at $\vec q=2(\vec k_0\pm \vec k_e)$,
and the linear loss of scattered field in the Maxwell-Bloch equation.
Within the time scale of interest, losses in $N_2$ due to its
coupling to third order side modes are negligible.
The motional ground state-condensate node, is coupled only to the two
first order side modes thus no dissipative terms appear in the
equation for $N_0$. In view of the one
to one correspondence between the number of atoms in the side
modes and their corresponding number of scattered photons, we will
further assume $\kappa={\cal L}$. In Ref. \cite{inouye} it
was estimated that the decoherence time (the decay time of matter
wave interference) was found to be $\sim 32-35$ ($\mu$s). In our
numerical simulations, we will thus take dephasing rate varies in
the range $\gamma_\bot\sim (2\pi)\, 0.8-1.6\times10^4$ (Hz), while
${\cal L}\sim(2\pi)0.5\times10^4$(Hz). The
coherent coupling parameter $g$ depends on the pump laser power.
Ref. \cite{inouye} reports a typical Rayleigh scattering rate
$\sim (2\pi)\, 7-700$ (Hz) at pump intensities $\sim 1-100$
(mW/cm$^2$). We thus take $\gamma/(2\pi)=$5-15 (Hz). The initial
condensate number is chosen to be $N_0=4\times10^6$ \cite{inouye}
for all numerical simulations.

In Figures \ref{fig6}-\ref{fig9},
$\gamma_\bot=(2\pi)\, 1.6\times 10^4$ (Hz) is used, while
$\gamma_\bot=(2\pi)\, 8\times 10^3$ (Hz)
is used in Figs. \ref{fig10}-\ref{fig13}.
The depletion of condensate atoms is shown in
Figs. \ref{fig6} for several different choices
of $\gamma/(2\pi)=$5.1, 6.7, 8.3, and 10.7 (Hz) with
larger coupling rates correspond to faster decaying curves.
The effect of $\gamma_\bot$ on the condensate decay
can be obtained from a comparison of Fig. \ref{fig6}
with Fig. \ref{fig10}, in which only
two curves with $\gamma/(2\pi)=5.1$ and 10.7 (Hz) are displayed.
Looking at these two curves, we can identify four separate dynamical regimes.
First, there is the linear regime where the condensate
can be considered as undepleted.
Second, there is the super-radiant regime
where a fast decay of condensate atom number occurs.
The third regime is a transient one where
oscillatory behavior is seen. Finally we have the
fourth regime where saturation sets in and
the condensate decays slowly.
Figs. \ref{fig6} and \ref{fig10} show
that oscillations are more prominent at
smaller decoherence rate and higher pump laser power.
In this case, the linear regime is shortened and
super-radiant decay becomes faster.

Figures \ref{fig7} and \ref{fig8} display matter wave
amplifications for the first and second atomic side modes
respectively for the same parameter set as used in Fig.
\ref{fig6}; They should be compared with Figs. \ref{fig11} and
\ref{fig12} where the same parameter set as in Fig. \ref{fig10}
are used . We note that shorter and more intense super-radiant
pulses are obtained at higher pump laser powers. Figs. \ref{fig9}
and \ref{fig13} represent the temporal evolution of scattered
light intensities for the two sets of parameters used for Figs.
\ref{fig6} and \ref{fig10} respectively. They are seen to increase
sharply but to decay more slowly. At higher laser powers double
peak (shoulder) structures are seen while at lower laser powers
the pulse shape becomes more asymmetric, broad and has single
peak. With lower decoherence rate the double-peak structure is
more prominent, and the pulse becomes more intense and shorter.
Additional peaks in the rings are also observed in tail regions.
Physically we assign the double-peak structure as due to
sequential (cascading) super-radiant scattering.

Overall we find that the high order decorrelation approximation
produces simulation results capable of
capturing detailed dynamics of the
super-radiant scattering from trapped condensates \cite{inouye}.
Apart from the unavoidable choice of introducing and adjusting
phenomenological coupling constants and various decay and
decoherence rates, all reported experimental observations
can be interpreted based on our model \cite{inouye}.
We also find that $Z$ and $W$ terms are almost negligible
in affecting system dynamics, presumably because they are already
of a higher order as compared with $X$ and $Y$. Further research
into this point is needed for a complete characterization of
quantum states of scattered photons.

For completeness, we finally discuss solutions to
Eq. (\ref{eqp}) using the Markov approximation. We formally
integrate the Heisenberg operator equation for $b_{\vec k}(t)$
in terms its initial condition $b_{\vec k}(0)$ and a radiation
reaction term related to emitted field from atoms.
The formal solution is then substituted into
atomic operator equations and the standard Markov approximation
made such that the radiation
reaction on atoms become instantaneous. The resulting Langevin
equations for atomic operators now contain quantum noise terms \cite{moore}
due to $b_{\vec k}(0)$ and $b_{\vec k}^\dag (0)$. Averaging
over this quantum noise reservoir and taking care of operator
ordering, we obtain the following equations for averaged
quantities (again neglecting $\langle.\rangle$)
\begin{eqnarray}\label{markov}
\frac{d}{dt}N_0&=&-\gamma_{M}N_0(1+N_1),\nonumber\\
\frac{d}{dt}N_1&=&\gamma_{M}[N_0(1+N_1)-N_2(1+N_2)]-{\cal L}N_{1},\nonumber\\
\frac{d}{dt}N_0&=&\gamma_{M}N_1(1+N_2),
\label{eqm}
\end{eqnarray}
with $\gamma_{M}$ the Markovian coupling constant
$\propto\gamma^2$. It also depends on reservoir
noise spectra width as well as a shape factor
of the condensate.
The above procedure is similar to what is used in
obtaining Eqs. (\ref{fdot}),
(\ref{bdot}), (\ref{fsmall}), (\ref{fl}), and (\ref{fav}).
$\gamma_M$ can be estimated from the Rayleigh scattering rate
$R$ to be
$\gamma_M=3R\Omega/8\pi$ (where typically in the MIT experimental
set up \cite{inouye}, $\Omega\sim 2\times 10^{-4}$).
We again will use $N_0=4\times 10^6$ and the new
phenomenological loss rate
[due to scattering into nodes at $\vec q=2 (\vec k_0\pm \vec k_e)$
in Fig. \ref{fig5}]
is chosen to be ${\cal L}=(2\pi)\, 4.71\times10^3$ (Hz).

The results for condensate and side mode populations
from Markovian dynamics Eq. (\ref{eqm}) are presented
in Figs. \ref{fig14}-\ref{fig16}. The four sets of
curves are gain respectively for
$\gamma_M/(2\pi)=3.18\times 10^{-3}$, $6.37\times 10^{-3}$,
$3.18\times 10^{-2}$, and $1.2\times 10^{-1}$ (Hz),
with larger $\gamma_M$ values
correspond to faster decaying condensates.
We see that the
transient regime with oscillations is visibly
absent, while previous results from the decorrelation
approximation [Eqs. (\ref{nm1}) and (\ref{nm2})]
correctly captures this essential experimental feature.
In addition, we find the scattered intensity
displays a two pulse structure in the Markov
approximation, rather than the ringing shoulder
structure discussed earlier.
We thus come to the conclusion that a simple Markov
approximation is incapable of describing
dynamic processes of the super-radiant BEC system.
In the Markovian limit, the dynamical behavior of
scattered intensity follows $I\sim N_0(1+N_1)+N_2(1+N_2)$.
By incorporating quantum fluctuation and
choosing appropriate initial conditions, it may be possible to
find averaged intensity distribution from
repeated simulations. The study in
Ref. \cite{moore} presented results from a collection of
single-shot simulations.

\section{CONCLUSIONS}
In summary, we have presented a thorough investigation of the
super-radiant light scattering from trapped Bose condensates. We
have shown that depending on gas sample conditions: {\it e.g.}
density, geometrical shapes, temperatures, and pump field
characteristics, the scattered light can exhibit either
oscillatory or exponential gain behaviors. We have presented a new
atomic side mode formulation that allows for a useful and simple
framework to analyze problems related to scattering from trapped
particles in momentum space. A cascading structure in the side
mode lattice is presented that allows for identification of
sequential super-radiant pulse dynamics. Our investigations also
point to eminent correlations between photons from the two end
firing modes similar to the two cascading photons from a single
atom \cite{jd}. Interestingly, atomic side modes connected through
multiple scattering are highly correlated because of their
non-commuting
algebra. Even though we simplified that aspect of them in
accordance with the current experimental designs, future
experiments might as well take this as an advantage of generating
strongly correlated high flux of photons. We are currently
investigating prospects of a super-radiant source for correlated
or entangled photons from our model system. Our study also
clarifies similarities and differences of the recent MIT
super-radiant experiment from trapped atomic BEC with earlier
studies of hot gas superfluorescence. We have compared theoretical
approaches based a Markov approximation with a non-Markovian
description of the experimental light scattering observation. We
find that the occurrence of multiple peaks in super-radiant pulse
reflects the generally non-Markovian nature of our system. Our
studies also sheds light on the role of dephasing in a coherent
quantum system \cite{kumar2}. Finally we note that a simple
modification in atomic side mode definition allows for studies of
super-radiant emission from a quantum degenerate trapped fermi
gas.

\section{ACKNOWLEDGMENT}
We thank Dr. C. Raman for helpful discussions.
This work is supported by the ONR grant No. 14-97-1-0633 and by the
NSF grant No. PHY-9722410.

\begin{figure}[b]
\centerline{\epsfig{file=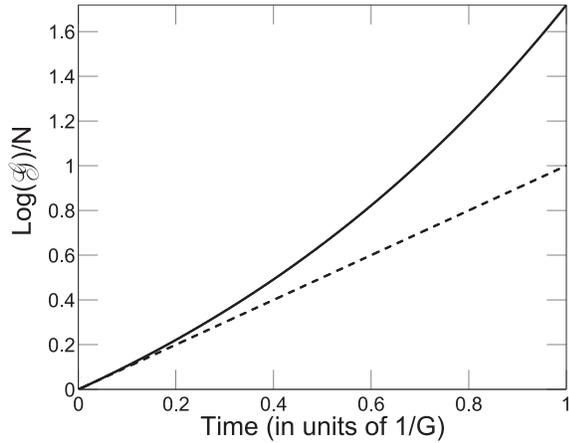,width=3.in}}
\caption{Comparison of different growth curves ${\cal G}_{\vec
q}(t)$ for an initial motional ground state condensate with
different statistics. Dashed line is for the Fock state and solid
line is for the coherent state. Semi-classical treatment also
predicts a growth same with the Fock state result} \label{fig4}
\end{figure}

\begin{figure}[b]
\centerline{\epsfig{file=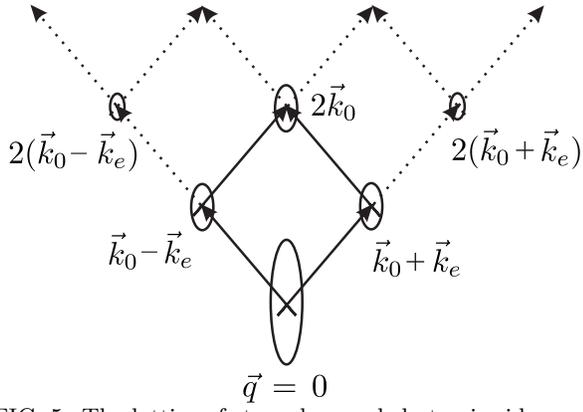,width=3.in}}
\caption{The lattice of strongly coupled atomic side modes. Solid lines show
the diamond shaped unit cell with cascade dynamics.} \label{fig5}
\end{figure}

\begin{figure}[t]
\centerline{\epsfig{file=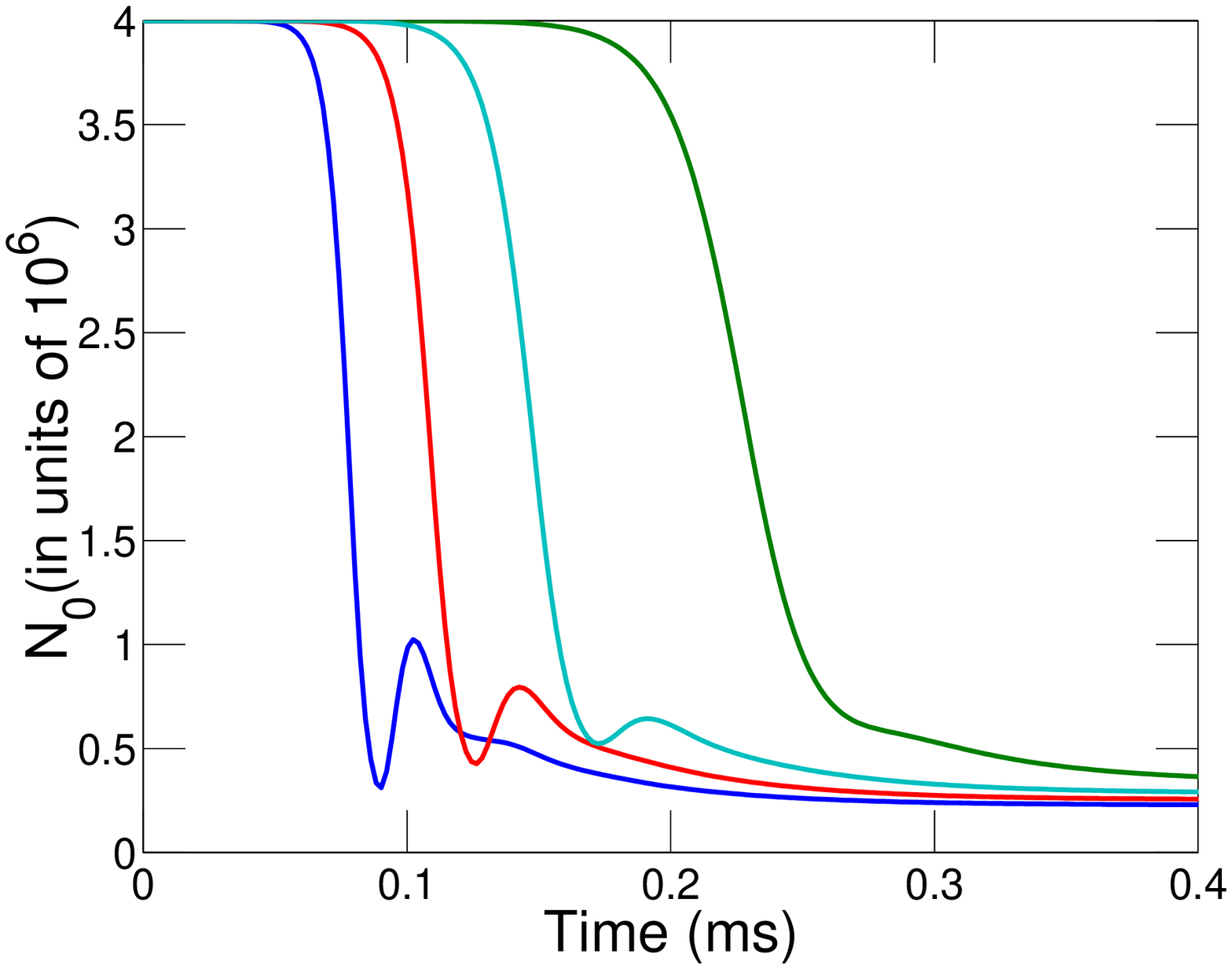,width=3.in}}
  \caption{The decay of condensate population.}
  \label{fig6}
\end{figure}

\begin{figure}[b]
\centerline{\epsfig{file=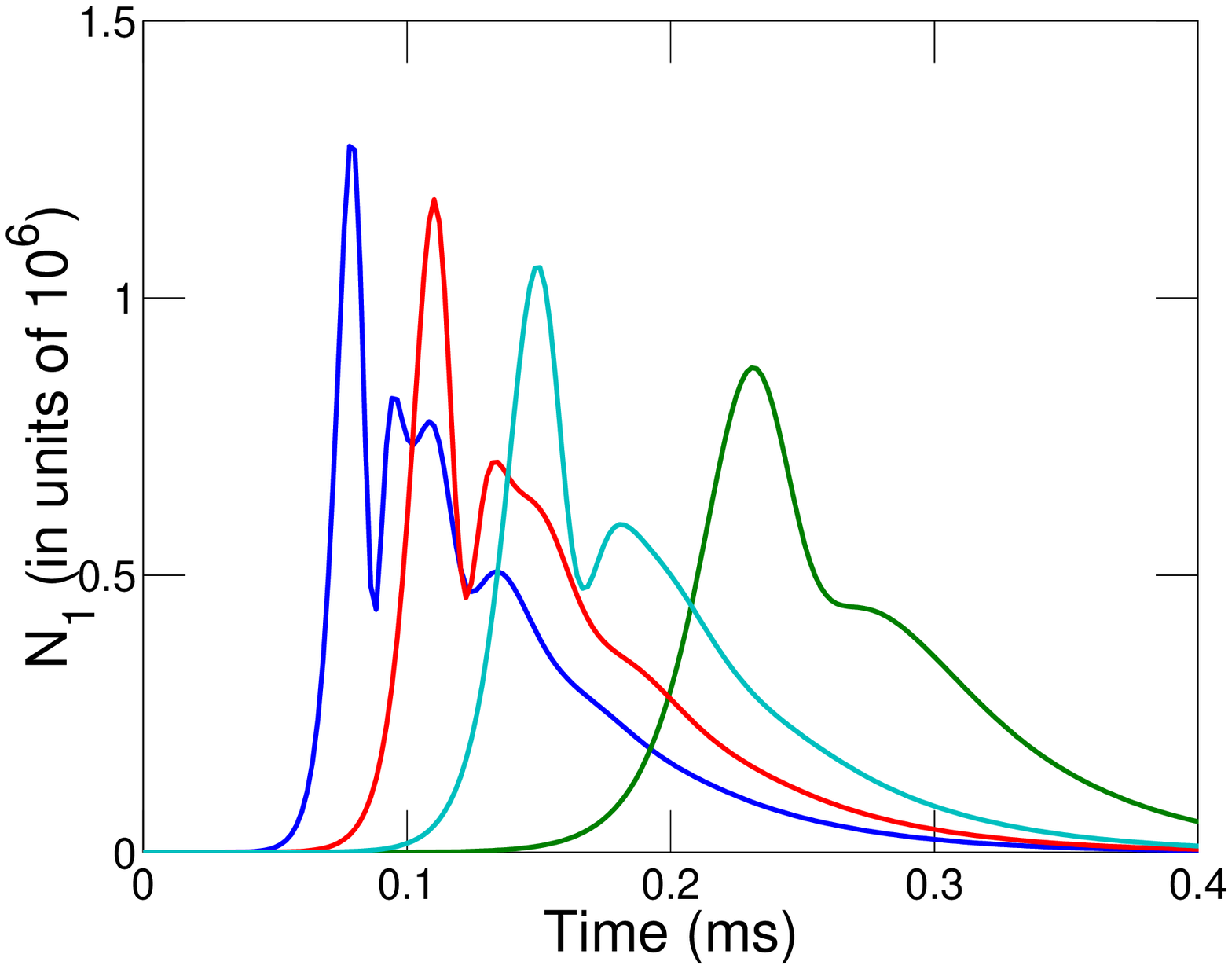,width=3.in}}
  \caption{The atomic population dynamics for the first side mode(s).}
  \label{fig7}
\end{figure}

\begin{figure}[t]
  \centerline{\epsfig{file=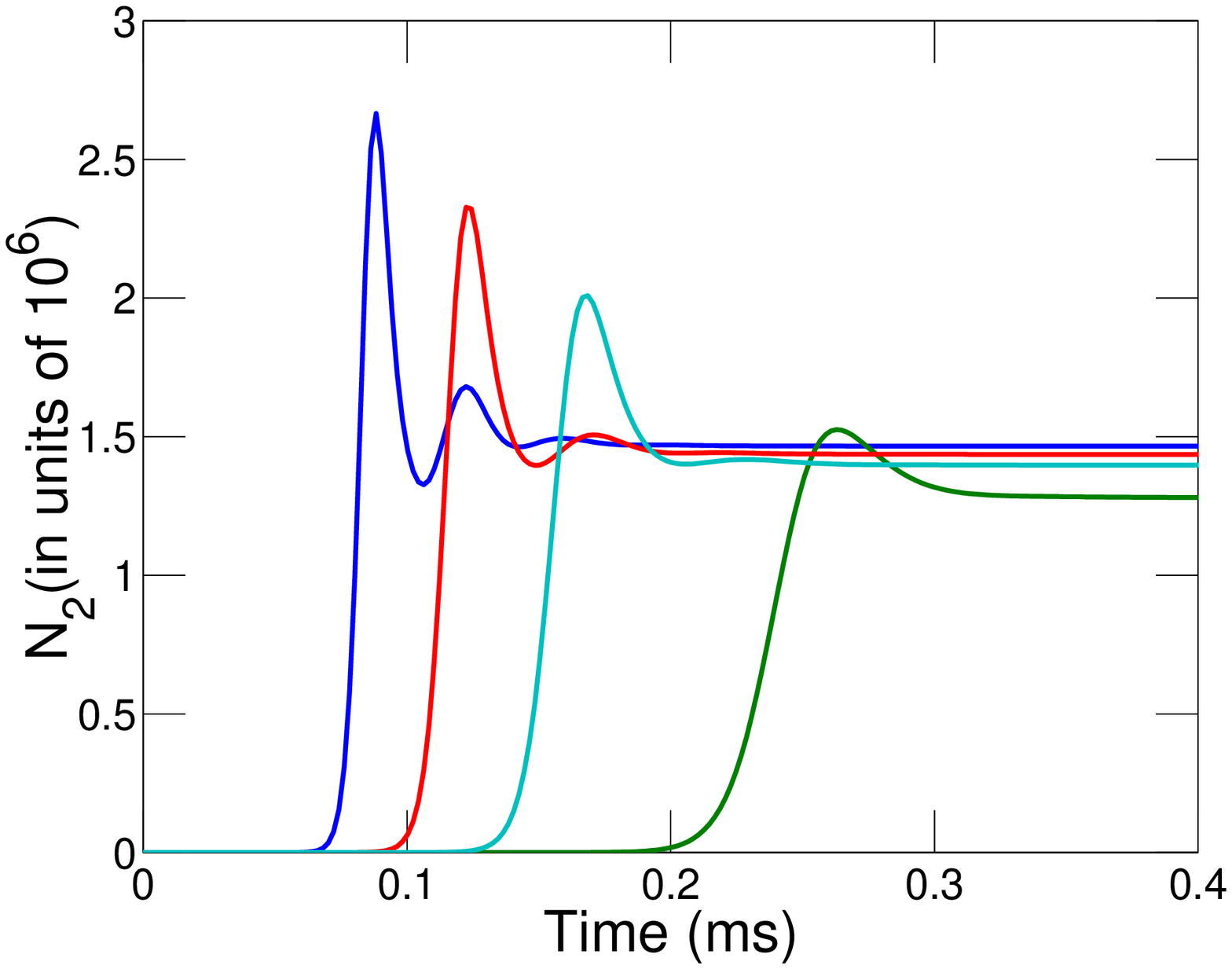,width=3.in}}
  \caption{The same as in Fig. 6 but for the second side mode.}
  \label{fig8}
\end{figure}

\begin{figure}[b]
  \centerline{\epsfig{file=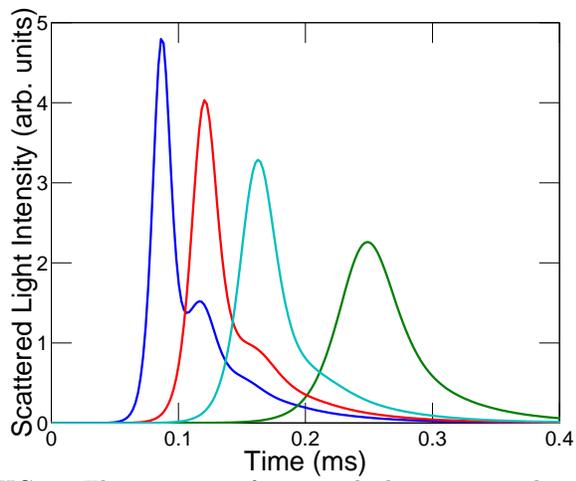,width=3.in}}
  \caption{The intensity of scattered photons around one of the
  end firing modes.}
  \label{fig9}
\end{figure}

\begin{figure}[t]
\centerline{\epsfig{file=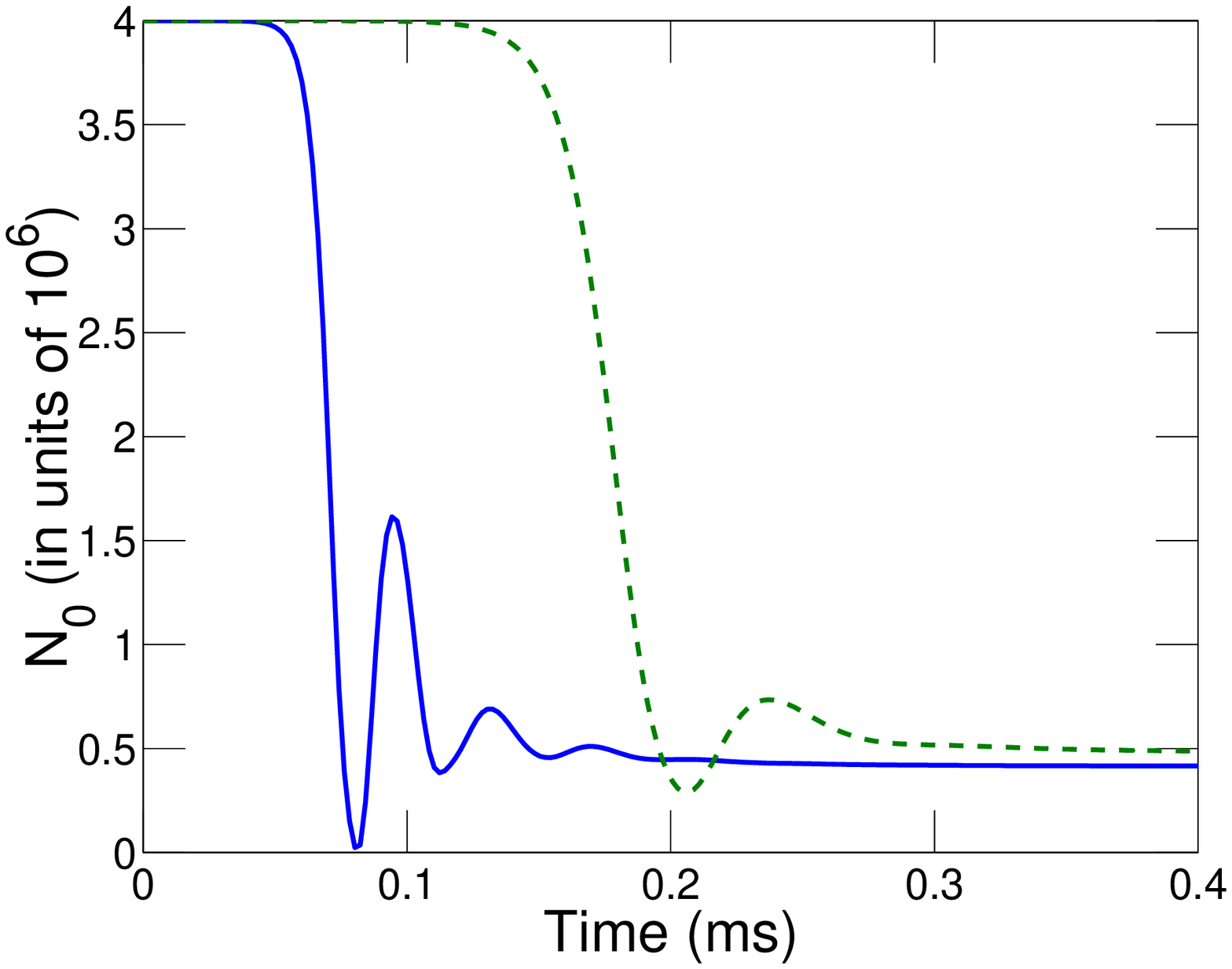,width=3.in}}
  \caption{The decay of condensate population.}
  \label{fig10}
\end{figure}

\begin{figure}
\centerline{\epsfig{file=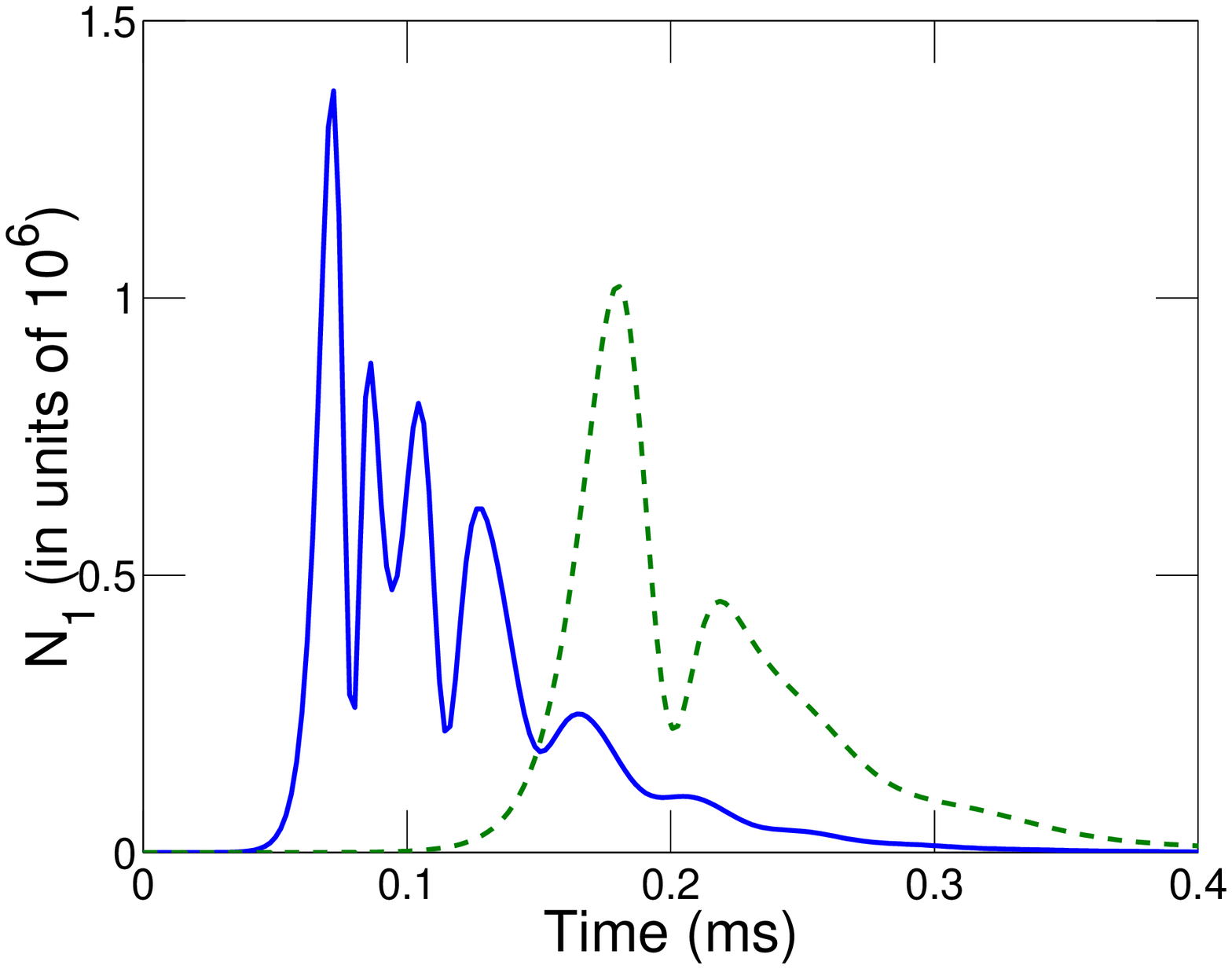,width=3.in}}
  \caption{The atomic population dynamics of the first side mode(s).}
  \label{fig11}
\end{figure}

\begin{figure}
\centerline{\epsfig{file=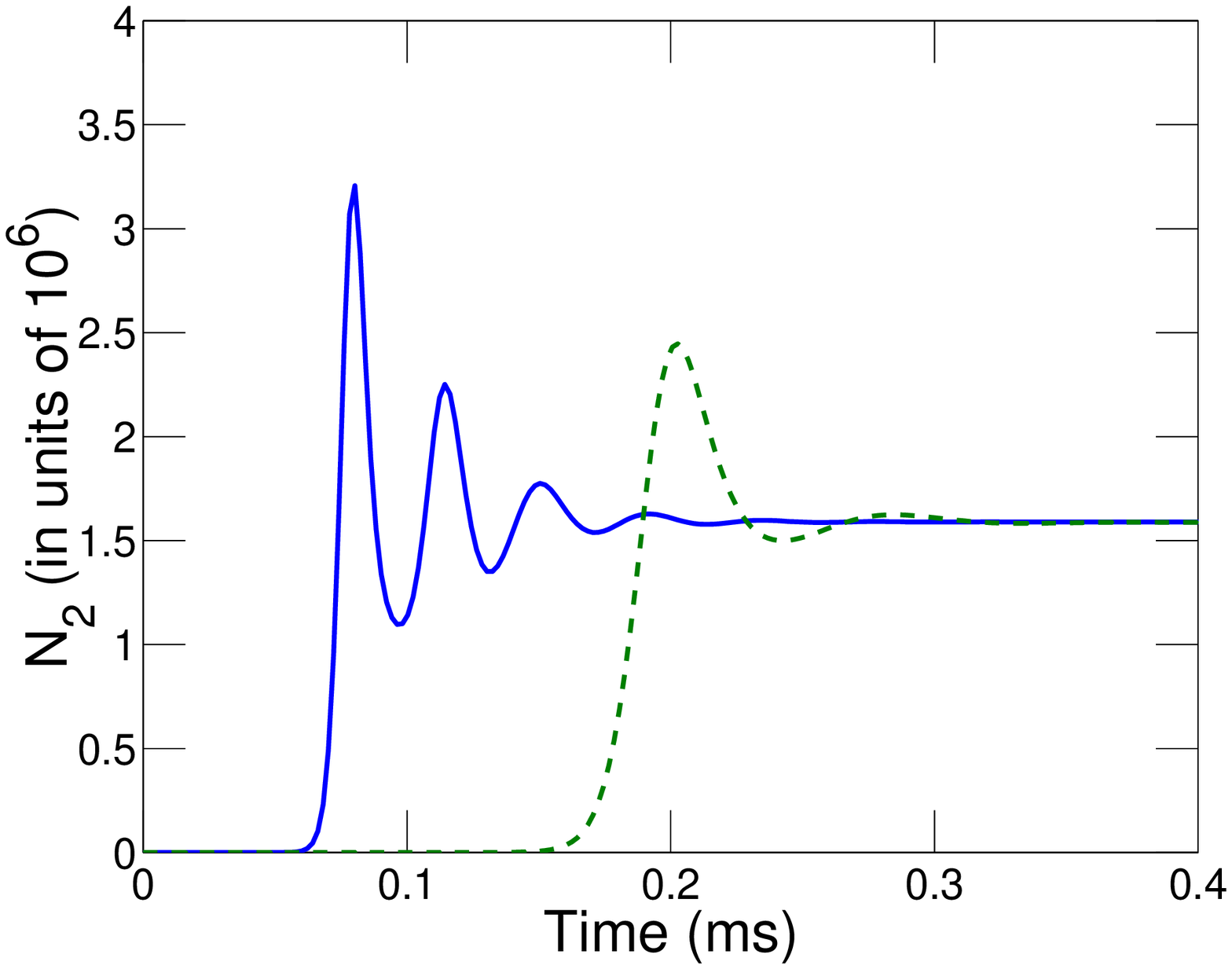,width=3.in}}
 \caption{The same as in Fig. 8 but for the second side mode.}
  \label{fig12}
\end{figure}

\begin{figure}
\centerline{\epsfig{file=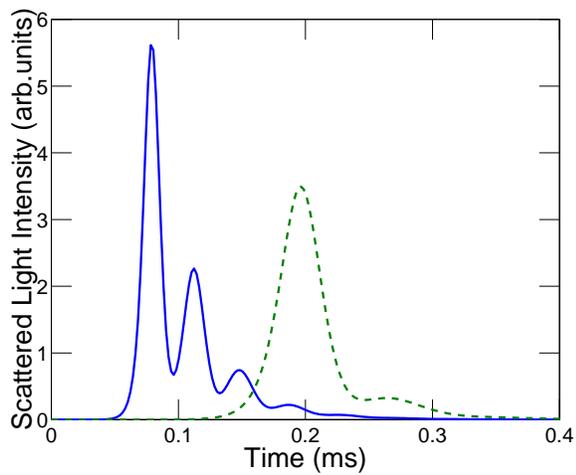,width=3.in}}
  \caption{The intensity of scattered field intensity around
  one of the end firing modes.}
  \label{fig13}
\end{figure}

\begin{figure}
\centerline{\epsfig{file=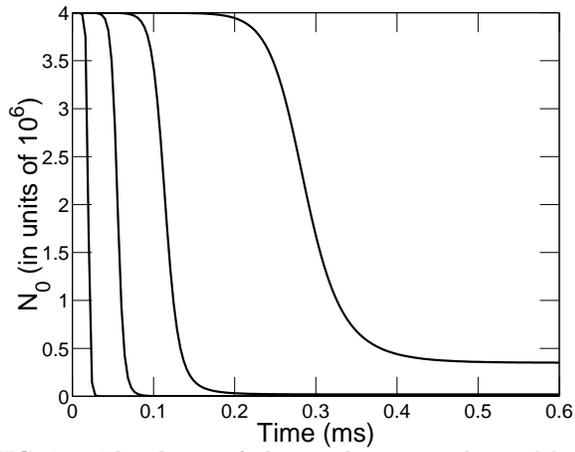,width=3.in}}
  \caption{The decay of the condensate
  within a Markov approximation for scattered light.}
  \label{fig14}
\end{figure}

\begin{figure}
\centerline{\epsfig{file=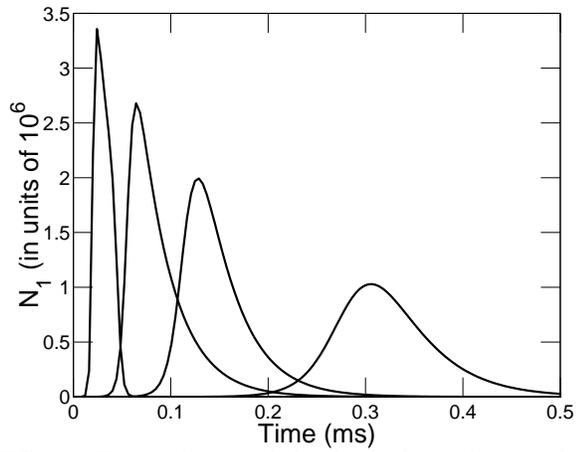,width=3.in}}
  \caption{The evolution of the first side mode(s) within a
  Markov approximation for scattered light.}
  \label{fig15}
\end{figure}

\begin{figure}
\centerline{\epsfig{file=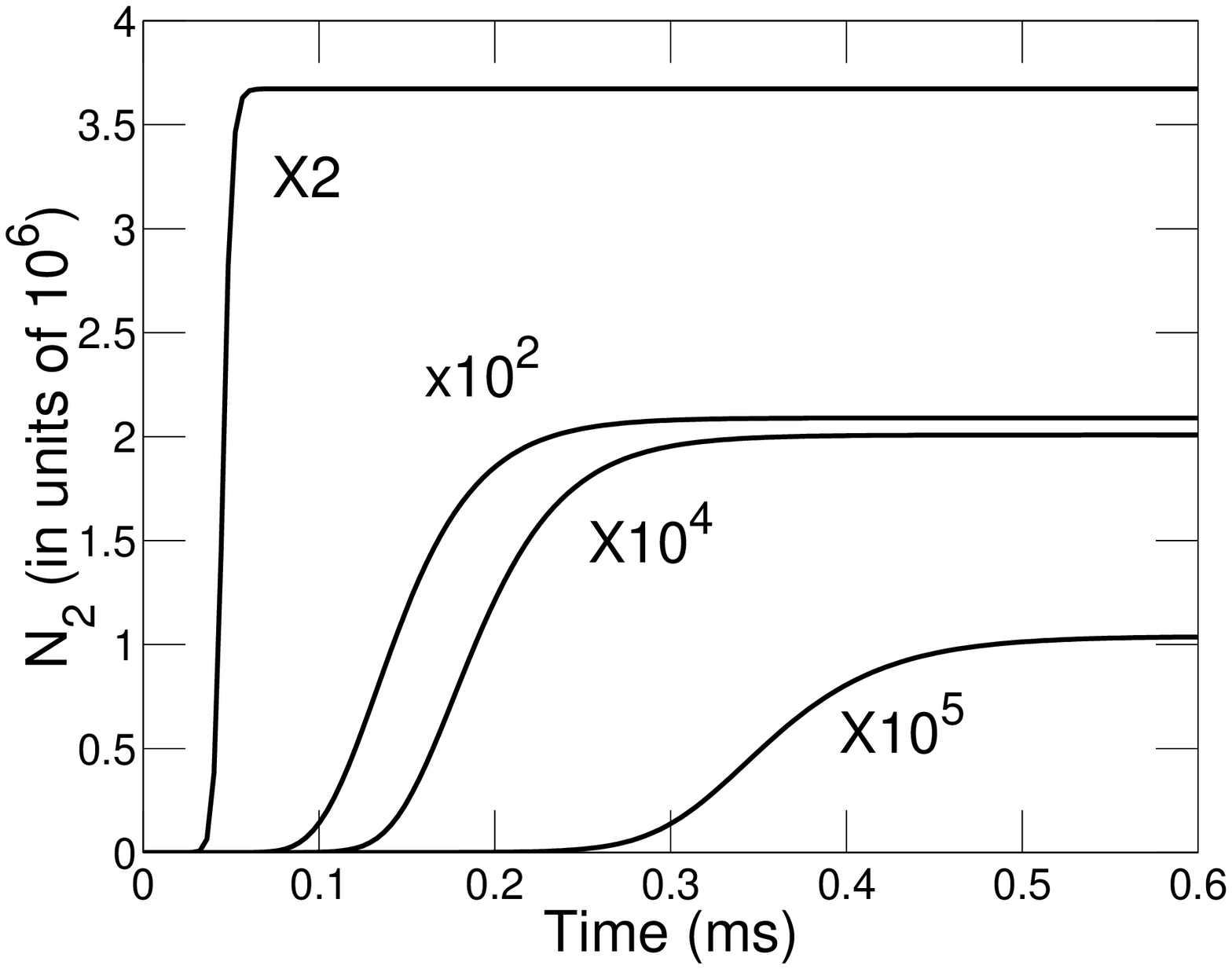,width=3.in}}
  \caption{The growth of second side mode within a
  Markov approximation for scattered light. Curves are
  magnified for better view.}
  \label{fig16}
\end{figure}


\begin{references}
\bibitem{anderson}M. H. Anderson, J. R. Ensher, M. R. Matthews,
C. E. Wieman, and E. A. Cornell, Science {\bf 269}, 198 (1995);
K. B. Davis, M. -O. Mewes, M. R. Andrews, N. J. van Druten,
D. S. Durfee, D. M. Kurn, and W. Ketterle, Phys. Rev. Lett. {\bf 75},
3969 (1995); C. C. Bradley, C. A. Sackett, J. J. Tollett, and R. G.
Hulet, {\it ibid} {\bf 75}, 1687 (1995);  {\bf 79}, 1170
(1997).

\bibitem{jila}E. A. Burt, R. W. Ghrist, C. J. Myatt, M. J. Holland,
E. A. Cornell, and C. E. Wieman, Phys. Rev. Lett. {\bf 79}, 337 (1997);
M. R. Andrews, C. G. Townsend, H.-J. Miesner, D. S. Durfee, D. M. Kurn,
W. Ketterle, Science {\bf 275}, 637 (1997).

\bibitem{deng} L. Deng, E. W. Hagley, J. Wen, M. Trippenbach, Y. Band, P. S. Julienne,
J. E. Simsarian, K. Helmerson, S. L. Rolston, and W. D. Phillips,
Nature {\bf 398}, 218 (1999).

\bibitem{inouye} S. Inouye, A. P. Chikkatur, D. M. stamper-Kurn, J. Stenger,
D. E. Pritchard and W. Ketterle, Science {\bf 285}, 571 (1999).

\bibitem{inouye2} S. Inouye, T. Pfau, S. Gupta, A. P. Chikkatur, A. G\"{o}rlitz,
D. E. Pritchard, and W. Ketterle, Nature {\bf 402}, 641 (1999).

\bibitem{moore} M. G. Moore and P. Meystre, Phys. Rev. Lett. {\bf 83},
5202 (1999).

\bibitem{elena} E. V. Goldstein and P. Meystre, Phys. Rev. A {\bf 59},
3896 (1999).

\bibitem{juha} J. Javanainen and J. Ruostekoski, Phys. Rev. A {\bf 52},
3033 (1995).

\bibitem{you}L. You, M. Lewenstein, and J. Cooper, Phys. Rev. A {\bf 51},
4712 (1995).

\bibitem{moore-zobay} M. G. Moore, O. Zobay, and P. Meystre,
Phys. Rev. A {\bf 60}, 1491 (1999).

\bibitem{stenger} J. Stenger, S. Inouye, A. P. Chikkatur, D. M. Stamper-Kurn,
D. E. Pritchard, and W. Ketterle, Phys. Rev. Lett. {\bf 23}, 4569
(1999).

\bibitem{kozuma} M. Kozuma, L. Deng, E. W. Hagley, J. Wen, R. Lutvak,
K. Helmerson, S. L. Rolston, and W. D. Phillips, Phys. Rev. Lett.
{\bf 82}, 871 (1999).

\bibitem{po}J. J. Hopfield, Phys. Rev. {\bf 112}, 1555 (1958).

\bibitem{bonifacio} R. Bonifacio and L. A. Lugiato, Phys. Rev. A
{\bf 11}, 1507 (1975).

\bibitem{gardiner}C. W. Gardiner, Phys. Rev. A {\bf 56}, 1414 (1997).

\bibitem{polder}D. Polder, M. F. H. Schuurmans, and Q. H. F. Vrehen,
Phys. Rev. A {\bf 19}, 1192 (1979).

\bibitem{skriban} N. Skribanowitz, I. P. Herman, J. C.
MacGillivray, and M. S. Feld, Phys. Rev. Lett. {\bf 30}, 309
(1973).

\bibitem{bonifacio2} R. Bonifacio, G. R. M. Robb,
and B. W. J. McNeil, Phys. Rev. A {\bf 56}, 912 (1997).

\bibitem{gibbs} H. M. Gibbs, Q. H. F. Vrehen, and H. M. J. Hikspoors,
Phys. Rev. Lett. {\bf 39}, 547 (1977).

\bibitem{bowden1}C. M. Bowden and C. C. Sung, Phys. Rev. A
{\bf 18}, 1558 (1978).

\bibitem{bowden2}C. M. Bowden and C. C. Sung, Phys. Rev. A
{\bf 20}, 2033 (1979).

\bibitem{you2}L.You, J. Cooper, and M. Trippenbach,
J. Opt. Soc. Am. B {\bf 8}, 1139 (1990).

\bibitem{rpa}
{\it Quantum Theory of the Optical and Electronic Properties of
Semiconductors}, H. Haug and S. W. Koch, (World Scientific,
Singapore, 1990).

\bibitem{new}
R. bonifacio and A.L.Lugiato, Phys. Rev. A {\bf 12}, 587 (1975);
AV Andreev, JETP {\bf 45}, 734 (1977);
AV Andreev, NA Enaki, and Yu A Ilinskii, JETP {\bf 60}, 229 (1984);
{\it Cooperative Effects in Optics: Superradiance and phase transitions},
AV Andreev, VI Emel'yanov, and Yu A Ilinskii,
(IOP publishing, Bristol, 1993).

\bibitem{watson} E. A. Watson, H. M. Gibbs, F. P. Mattar, M. Cormier, Y. Claude,
S. L. McCall, M. S. Feld, Phys. Rev. A {\bf 27}, 1427 (1983).

\bibitem{mattar} F. P. Mattar and C. M. Bowden, Phys. Rev. A {\bf
27}, 345 (1983).

\bibitem{mandel} L. Mandel and E. Wolf, {\it Optical coherence and
Quantum Optics}, (Cambridge University Press, New York, 1995),
p847.

\bibitem{heinzen} D. J. Heinzen, J. E. Thomas, and M.S. Feld, Phys.
Rev. Lett. {\bf 54}, 677 (1985).

\bibitem{kumar1} A. Kumarakrishnan and X. L. Han, Opt. Commun.
{\bf 109}, 348 (1994).

\bibitem{jd}A. Aspect, J. Dalibard, and G. Roger,
Phys. Rev. Lett. {\bf 49}, 1804 (1982).

\bibitem{kumar2} A. Kumarakrishnan and X. L. Han, Phys. Rev. A
{\bf 58}, 4153 (1998).

\end{references}
\end{document}